\documentclass[a4paper,11pt]{article}
\usepackage{amssymb,amsmath,amsfonts,placeins,pbox,multirow,mathtools}
\usepackage{graphicx,rotate,color,slashed,cite,epstopdf,verbatim,url,multirow,booktabs}
\usepackage{epstopdf}
\usepackage{subfig}
\usepackage{bm}
\usepackage{graphicx,caption}
\usepackage{array}
\usepackage[T1]{fontenc}
 
\usepackage[a4paper, left=2cm, right=2cm, top=2.5cm, bottom=2.5cm]{geometry}
\usepackage[colorlinks=true,
            linkcolor=magenta,
            urlcolor=blue,
            citecolor=blue]{hyperref}

\numberwithin{equation}{section}

\def\ssbh#1//#2//{\ensuremath{\xrightarrow [\substack{#2}]
    {\parbox{3cm}{\hfil $\scriptstyle \langle #1 \rangle$ \hfil}}}}

\let\CapTion=\caption
\def\caption#1{\CapTion{\em #1}}

\usepackage[capitalise]{cleveref}
\crefname{section}{Sec.}{Secs.}
\crefname{table}{Tab.}{Tabs.}
\crefname{figure}{Fig.}{Figs.}
\crefname{equation}{Eq.}{Eqs.}
\crefname{appendix}{Appendix}{Appendix}

\title{\bf  Phase transition and gravitational waves in maximally symmetric composite Higgs model}

\author{\bf Avik Banerjee$^{a}$\thanks{avik.banerjee\_205@tifr.res.in}\,, Marco Merchand$^{b,c}$\thanks{marcomm@kth.se}\,, Ignacy Nałęcz$^{d}$\thanks{inalecz@fuw.edu.pl}
\\ \medskip 
{}$^{a}$\small\em Department of Theoretical Physics, Tata Institute of Fundamental Research,\\[-13pt] \small\em Homi Bhabha Road, Mumbai 400005, India\\ 
{}$^{b}$\small\em KTH Royal Institute of Technology, Department of Physics, SE-10691 Stockholm, Sweden\\ 
{}$^{c}$\small\em The Oskar Klein Centre for Cosmoparticle Physics, AlbaNova University Centre,\\[-8pt] \small\em SE-10691 Stockholm, Sweden \\ 
{}$^{d}$\small\em Institute of Theoretical Physics, Faculty of Physics, University of Warsaw, \\[-8pt] \small\em ul. Pasteura 5, 02-093 Warsaw, Poland
}
\date{}

\begin{document}

\maketitle

\begin{flushright}
\vspace{-11cm}
\begin{small}
TIFR/TH/24-7
\end{small}
\end{flushright}
\vspace{11cm}


\begin{abstract}

In this paper we study phase transitions in a maximally symmetric composite Higgs model with next-to-minimal coset, where a pseudoscalar singlet emerges alongside the Higgs doublet. The maximal symmetry guarantees  the finiteness of the radiatively generated scalar potential. We explore the scenario involving an explicit source of CP violation in the strong sector, which induces a $\mathbb{Z}_2$ asymmetric scalar potential, and consequently leads to nonzero vacuum expectation value for the singlet. Current experimental bounds from the LHC are imposed on the masses of the composite resonances, while the CP violating interactions of the pseudo Nambu-Goldstone bosons are tightly constrained from the measurements of the electric dipole moment of the electron. We compute the finite temperature corrections to the potential, incorporating the momentum-dependent form factors in the loop integrals to capture the effect of the strong dynamics. The impact of the resonances from the strong sector on the finite temperature potential are exponentially suppressed. The presence of explicit CP violation leads to strong first-order phase transition from a false vacuum to the electroweak vacuum where the pseudoscalar singlet has a non-zero vacuum expectation value. We illustrate that, as a result of such phase transitions, the production of potentially observable gravitational waves at future detectors will offer a complementary avenue to probe the composite Higgs models, distinct from collider experiments.

\end{abstract}


\bigskip

\newpage
{
  \hypersetup{linkcolor=black}
  \tableofcontents
}

\bigskip
\hrule
\bigskip

\section{Introduction}
\label{intro}

The origin of the Higgs boson as a composite pseudo Nambu-Goldstone boson (pNGB) emerging from a strongly coupled confining sector remains as an exciting possibility to address the electroweak (EW) hierarchy problem \cite{Kaplan:1983fs,Kaplan:1991dc,ArkaniHamed:2002qy,Contino:2003ve,Agashe:2004rs}. In this class of models the pNGB Higgs boson arises due to the spontaneous breaking of a global symmetry of the strong sector, see \cite{Contino:2010rs,Panico:2015jxa} for comprehensive reviews. Interestingly, majority of the models featuring a 4D confining gauge theory with fermionic matter content (known as hyperquarks) in the ultra-violet (UV) yield new pNGBs in addition to the usual Higgs doublet \cite{Barnard:2013zea,Ferretti:2013kya}. The large value of the Yukawa coupling of the top quark is explained in these scenarios by requiring that the top quark is partially composite \cite{Kaplan:1991dc,Contino:2004vy}.  These models are also studied in the holographic setup \cite{Contino:2003ve,Contino:2004vy,Erdmenger:2020lvq,Erdmenger:2020flu} as well as on the lattice \cite{Bennett:2023wjw,Ayyar:2017qdf}.

Major predictions of the models with pNGB Higgs and partially composite top quark are the existence of spin-1 resonances and colored fermions (top-partners) around the TeV scale, comparatively lighter pNGBs in addition to the Higgs boson, and modifications of the Higgs couplings with other Standard Model (SM) particles. Needless to say, the non-observation of any signature for physics beyond the Standard Model (BSM) at the Large Hadron Collider (LHC) squeezes the parameter space of these models, compelling us to allow for some degree of fine-tuning. 

In the next decade, the LHC will enter into its high luminosity phase (HL-LHC), however the centre of mass energy will not change significantly. Thus, HL-LHC will probe the couplings of the Higgs boson up to a few percent level, and potentially measure its self-coupling. In view of this situation, it is worthwhile to consider complementary probes of the BSM physics beyond the collider experiments. In this context, the potential detection of stochastic gravitational waves (GWs) resulting from a first-order phase transition (FOPT) in the early universe by upcoming interferometer experiments, such as LISA, AEDGE, AION \cite{2017arXiv170200786A,AEDGE:2019nxb, Badurina:2019hst}, among others, presents a compelling alternative. In fact, such a future detection may offer deep insight into the structure of the Higgs potential.

The pNGB potential is generated when the global symmetries of the strong sector are explicitly broken by external sources, such as a mass term for the hyperquarks, or the gauge and Yukawa interactions of the pNGBs. Under the minimal Higgs potential (MHP) hypothesis \cite{Marzocca:2012zn}, the dominant contributions to the pNGB Higgs potential are induced in the infra-red (IR) by the gauge and Yukawa interactions through the radiative Coleman-Weinberg mechanism. Crucially, the contribution from the top quark is essential to trigger electroweak symmetry breaking by the vacuum misalignment mechanism \cite{Banerjee:2023ipb}. The hyperquark mass term also contributes to the potential, analogous to the quark mass contributions to the pion potential in quantum chromodynamics (QCD), if a 4D confining gauge theory is postulated in the UV.  Apart from the IR contributions, higher order operators in the chiral effective theory of the pNGBs can be constructed which explicitly break the global symmetries of the strong sector and provide UV contributions to the pNGB potential \cite{Marzocca:2012zn}.

The FOPT has been extensively studied in the composite Higgs models \cite{Espinosa:2011eu,Bruggisser:2018mus,Bruggisser:2018mrt,Bruggisser:2022rdm,vonHarling:2023dfl,Fujikura:2023fbi}, for example in the context of the next-to-minimal $SO(6)/SO(5)$ coset  \cite{Gripaios:2009pe,Redi:2012ha,Serra:2015xfa,Low:2015qep,Cai:2015bss,Arbey:2015exa,Niehoff:2016zso,Golterman:2017vdj,Alanne:2018wtp,Murnane:2018ynd,Bian:2019kmg,DeCurtis:2019rxl,Frigerio:2012uc,Marzocca:2014msa,Fonseca:2015gva,Kim:2016jbz,Espinosa:2011eu,Banerjee:2017qod}, which contains an additional SM singlet pseudoscalar.
The key motivation to consider this scenario arises from the fact that the modified scalar potential in the presence of additional pNGBs offers the possibility of realizing strong FOPT, which is well known to be absent in the SM. Additionally, the presence of extra scalars may provide new sources of CP violation which can aid in the generation of the baryon asymmetry via EW baryogenesis. 

However, the existing analyses in the next-to-minimal scenarios observe that the IR contributions to the pNGB potential, under the MHP hypothesis \cite{Marzocca:2012zn}, are not, in general, sufficient to yield strong FOPT. To evade this problem, several non-trivial extensions of the next-to-minimal model have been proposed, which can be broadly divided in two categories, i) where UV contributions to the pNGB potential are added \cite{Bian:2019kmg,Xie:2020bkl,Frandsen:2023vhu}, and ii) where additional particles such as a dilaton or top-partners with higher representations of $SO(6)$ are included \cite{Xie:2020bkl,Zhang:2024dgv}. It is worth noting that, in these analyses, the pseudoscalar singlet of the next-to-minimal coset does not receive a vacuum expectation value (vev) at low temperature.

In this paper, we consider a maximally symmetric composite Higgs model \cite{Csaki:2017cep,Csaki:2017jby,Serra:2017poj,Csaki:2018zzf,Dong:2020eqy,Blasi:2022hgi} based on the $SU(4)/Sp(4)\simeq SO(6)/SO(5)$ coset \cite{Barnard:2013zea,Ferretti:2013kya,Cacciapaglia:2014uja,Ferretti:2014qta,Agugliaro:2016clv,Galloway:2016fuo,Xing:2020uaf}, which arises from a confining gauge theory in the UV. We provide a novel way to compute an exact analytic form of the one-loop potential at both zero and finite temperature, taking into account the momentum-dependent form factors, which capture the information about the strong dynamics. We  reappraise the contributions of the IR potential to the strong FOPT in this model, while keeping the top-partners in the antisymmetric $\mathbf{6}$ of $SU(4)$. Below we describe some of the novel aspects of this study which have significant impact on the dynamics of FOPT and the production of GWs:

\begin{itemize}
    \item First of all, we study the pNGB potential for the next-to-minimal coset in the maximally symmetric limit. The maximal symmetry, introduced in \cite{Csaki:2017cep}, ensures the convergence of the one-loop Coleman-Weinberg potential for the pNGBs at zero temperature. Thus, unlike generic composite Higgs models, in the maximally symmetric cases the potential is fully calculable in terms of the parameters of the strong sector.

    \item As mentioned earlier, in the existing literature on $SU(4)/Sp(4)\simeq SO(6)/SO(5)$ coset, usually a $\mathbb{Z}_2$ symmetric potential is considered, under which the pseudoscalar singlet pNGB transforms oddly, $\eta \to -\eta$. As shown in \cite{Alanne:2018wtp}, this $\mathbb{Z}_2$ symmetry cannot be broken spontaneously by giving a vev to the singlet, unless the strong sector violates CP explicitly. We specifically consider the scenario  where CP is explicitly broken in the strong sector as well as in the top quark sector, owing to the presence of complex phases in the hyperquark mass matrix and the Yukawa interactions of the partially composite top quark. As a result, the pNGB potential explicitly breaks the $\mathbb{Z}_2$ symmetry generating a tadpole term for the singlet, and ultimately leading to a non-zero vev for it at zero temperature. 
   We study strong FOPT leading to a low temperature vacuum characterized by the non-zero vev of the singlet in the next-to-minimal coset with maximal symmetry, which to the best of our knowledge, marks the first exploration of its kind.

    \item We compute the thermal corrections to the pNGB potential using the imaginary time formalism. We incorporate the full momentum dependence of the form factors, which capture the strong dynamics, in the loop integrals. Despite the presence of non-trivial momentum dependence, the finite temperature potential appears to have the same functional form as the standard thermal functions derived by Dolan and Jackiw in \cite{Dolan:1973qd}. Furthermore, we illustrate an innovative method to exactly compute the one-loop potential at both zero and finite temperature, which yields a convergent analytic expression for the pNGB potential, owing to the existence of maximal symmetry.

    \item We demonstrate that, unlike the $\mathbb{Z}_2$ symmetric case, in the presence of an explicit CP violation in the strong sector, thermal evolution of the IR potential under the MHP hypothesis is sufficient to yield strong FOPT and consequently the production of GWs.
\end{itemize}

We systematically study the thermodynamics of the phase transition, identify the allowed region of parameter space which satisfies the constraints from the LHC, measurements of the electric dipole moment (EDM) of the electron and leads to strong FOPT. Further, with some motivated benchmark values of parameters we show that the GWs resulting from the strong FOPT lie within the sensitivity of the future detectors, such as LISA, AEDGE.

The paper is organized as follows: in \cref{msCHM} we introduce the maximally symmetric composite Higgs model and calculate the zero temperature potential for the pNGBs. The derivation of the finite temperature corrections to the potential is presented in section \cref{pot_thermal}. In \cref{FOPT} we discuss the constraints on the model parameters from the LHC, electron EDM measurement and the requirement of realizing a strong FOPT. We further show the GW spectra for some benchmark values of parameters in \cref{FOPT} before concluding in \cref{concl}.

\section{Maximally symmetric composite Higgs model}
\label{msCHM}

In this section we briefly review the concept of maximal symmetry in composite Higgs models \cite{Csaki:2017cep,Csaki:2017jby,Csaki:2018zzf}, focusing on a specific example based on the $SU(4)/Sp(4)$ coset. It is the minimal coset which can emerge from a 4D confining gauge theory with fermionic matter (hyperquarks), yielding five pNGBs at low energy including a $SU(2)$ doublet Higgs boson. 

A coset $G/H$ is called a symmetric coset space if the broken ($T^{\hat{a}}$) and unbroken ($T^a$) generators follow the commutation relations given below
\begin{align}
[T^a,T^b] = i f^{abc} T^c,\quad [T^{\hat{a}},T^{\hat{b}}] = i f^{\hat{a}\hat{b}c} T^c,\quad [T^a, T^{\hat{b}}] = i f^{a\hat{b}\hat{c}} T^{\hat{c}}.
\end{align}

Familiar examples of such spaces are $SU(N)/SO(N)$, $SU(2N)/Sp(2N)$, $SO(N)/SO(N-1)$ etc. The pNGBs ($\Pi\equiv \pi_{\hat{a}}T^{\hat{a}}$) are usually parameterized by $\Sigma \equiv \exp(i\sqrt{2}\Pi/f)$ which transforms non-linearly, $\Sigma\to g\Sigma h^{-1}(\Pi,g)$ where $g\in G$ and $h\in H$. In case of symmetric spaces, however, a parity operator $\epsilon_0$ exists such that the modified pNGB matrix $U=\Sigma^2\epsilon_0$ can be constructed  which transforms linearly as $U \to g U g^T$ (see \cite{Ferretti:2016upr,Banerjee:2022izw} for the construction of low energy Lagrangians). 
 
For the specific case of $SU(4)/Sp(4)$ coset, $\epsilon_0$ is associated with $Sp(4)$ invariant vacuum, defined as $\epsilon_0=i \sigma^3 \otimes \sigma^2$ and follows the relations
\begin{align}
T^a\epsilon_0+\epsilon_0 T^{aT} = 0, \quad T^{\hat{a}}\epsilon_0-\epsilon_0 T^{\hat{a}T} = 0.
\end{align}
In this case, the pNGBs transform as $\mathbf{5}$ of $Sp(4)$ which can be further decomposed under $Sp(4)\to SU(2)_L\times SU(2)_R \to SU(2)_L\times U(1)_Y$ as $\mathbf{5} \to (\mathbf{2,2})+(\mathbf{1,1}) \to \mathbf{2}_{\pm 1/2} (H) + \mathbf{1}_0 (\eta)
$. Here $H\equiv(H^+,H^0)$ denotes the usual Higgs doublet and $\eta$ is a SM gauge singlet. Detailed expressions for the generators and $\Sigma$ are given in the \cref{model_details}.

In the partial compositeness framework, the elementary quarks (we will primarily focus on the top quark due to its large Yukawa coupling with the Higgs boson) couple to the strong sector via linear mixing with composite top-partners. To construct the low energy Lagrangian after integrating out the heavy top-partners, elementary quarks are embedded into incomplete multiplets of the global symmetry $G$. For example, we embed both the left-handed doublet $q_L$ and the right-handed top quark $t_R$ in the anti-symmetric $\mathbf{6}$ of $SU(4)$. Thus, formally $G=SU(4)$ invariant effective Lagrangian can be written as\footnote{Following the usual practice, an additional factor of $U(1)_X$ is introduced to reproduce the correct hypercharge of the quarks, given by $Y=T^3_R+X$. The pNGBs are uncharged under this $U(1)_X$.}
\begin{align}
\nonumber
\mathcal{L}_{t} & = \bar{q}_L^i \slashed p \left[\Pi^L_0 (p^2) + \Pi^L_1 (p^2) {\rm tr} (A_L^i U^\dagger) {\rm tr} (U A_L^{i\dagger}) \right] q_L^i  + \bar{t}_R \slashed p \left[\Pi^R_0 (p^2) + \Pi^R_1 (p^2) {\rm tr} (A_R^\dagger U) {\rm tr} (U^\dagger A_R) \right] t_R \\
& + \bar{q}_L^i \left[\Pi^{LR}_1(p^2) {\rm tr} (A_L^i U^\dagger) {\rm tr} (U^\dagger A_R) +\Pi^{LR}_2(p^2) {\rm tr} (A_L^i U^* A_R U^\dagger) \right] t_R + {\rm h.c.}, 
\label{top_lag}
\end{align}
where $A^i_{L}$, ($i=t,b$) and $A_R$ denote the spurions of $SU(4)$ shown in \cref{form__factors} (see also ref. \cite{Banerjee:2022izw}). The momentum-dependent form factors $\Pi^{L,R}_{0,1}$ and $\Pi^{LR}_{1,2}$ encode the effect of the strong dynamics, see \cref{form__factors} for explicit expressions in terms of parameters of the composite resonances. 

If the interaction terms with $U$ are turned off, the above Lagrangian enjoys a chiral $G_L\times G_R$ symmetry due to the spurionic transformations $A_{L} \to g_{L}A_{L}g^T_{L}$ and $A_{R} \to g_{R}A_{R}g^T_{R}$, where $g_{L,R}\in G_{L,R}$. Although, the pNGB interactions in general break this symmetry, the term proportional to $\Pi^{LR}_2$ preserves a diagonal subgroup $G_D\supset G_L\times G_R$, defined by $g_L U g_R^T=U$. Thus, in the limit $\Pi^{L,R}_1=\Pi^{LR}_1=0$, the Lagrangian enjoys this remnant symmetry $G_D$, known as the \emph{maximal symmetry}. The UV origin of the maximal symmetry can provide some justification for the above conditions on the form factors, as explained in \cite{Csaki:2017cep}. Essentially this implies the following specific relations between the couplings and the masses of the lightest resonances in the UV
\begin{align}
\lambda^{L,R}_1=\lambda^{L,R}_5 \equiv \lambda^{L,R}, \qquad {\rm and} \qquad M_1=M_5\,,
\label{eq:maxsym}
\end{align}
where $M_5$ and $M_1$ denote the masses of the composite resonances, arising from a $\mathbf{6}$ of $SU(4)$ and transform as 5-plet and singlet under $Sp(4)$, respectively, while $\lambda$'s denote their corresponding couplings with the elementary quarks \emph{a la} partial compositeness. The existence of a maximal symmetry ensures that the top quark contribution to the one-loop Coleman-Weinberg potential for the pNGBs is finite.

The situation in the gauge sector of a symmetric coset is more involved, as discussed in \cite{Csaki:2017cep}. 
Although maximal symmetry is not fully realized in the gauge sector, the introduction of certain sum rules, inspired by the Weinberg sum rules in QCD, on the masses and decay constants of spin-1 resonances effectively render the one-loop pNGB potential finite (see \cite{Csaki:2017cep} for a detailed discussion). The effective low energy Lagrangian in the gauge sector, obtained after integrating out the spin-1 composite resonances, is given by\cite{Contino:2010rs}
\begin{align}
    \mathcal{L}_{g} = \frac{P_T^{\mu\nu}}{2}\left[g^2\Pi^W_0(p^2)W^a_{\mu}W^a_{\nu}+g^{\prime 2}\Pi^B_0(p^2)B_{\mu}B_{\nu} + \frac{\Pi_1(p^2)}{4}{\rm tr}((A_\mu U + U A_\mu^T)(A_\nu U + U A_\nu^T)^\dagger)\right],
    \label{gauge_lag}
\end{align}
where $A_\mu \equiv g W^a_\mu T^a_L +g^\prime B_\mu T^3_R$, $P^{\mu\nu}_T \equiv (\eta^{\mu\nu}-p^\mu p^\nu/p^2)$ is the transverse projector, and the form factors are given in \cref{form__factors}. The following sum rules are adopted to ensure convergence of the one-loop potential
\begin{align}
    2(f_\rho^2-f_a^2) = f^2\,, \qquad f_\rho^2 m_\rho^2 - f_a^2 m_a^2 =0\,, \qquad {\rm and} \qquad m_\rho \ne m_a\,,
    \label{sumrules}
\end{align}
where $f$ is decay constant of the Higgs boson, while $f_\rho$, and $f_a$ denote the decay constants of the lightest vector and axial-vector resonances $\rho_\mu$ and $a_\mu$, with masses $m_\rho$ and $m_a$, respectively. In the next subsection, we demonstrate the finiteness of the one-loop Coleman-Weinberg pNGB potential imposing maximal symmetry and using the leading-order approximation, while in \cref{pot_thermal}, we extend the calculation to full generality.

\subsection{pNGB potential at zero temperature}
\label{pot_zero}

At zero temperature, the pNGB potential arises due to explicit breaking of the global symmetry of the strong sector and receives three major contributions
\begin{align}
    V_{1\text{-loop}}(h,\eta) = V_t + V_g + V_H,
    \label{eq:V_tot}
\end{align}
where $V_t$ and $V_g$ are the one-loop contributions from the top quark, and the weak gauge bosons, respectively, while $V_H$ denotes the contribution from the hyperquark masses.  
Now we examine each of these contributions, followed by a comprehensive summary of the entire potential.

\subsubsection*{Top quark contribution}
\label{pot_zero_top}

The top quark contribution to the  Coleman-Weinberg potential can be calculated from the Lagrangian \eqref{top_lag} as
\begin{align}
    V_t = -2N_c \int \frac{d^4p}{(2\pi)^4} \log\left[1+\frac{|\Pi^{LR}_2(p^2)|^2}{p^2(1+\Pi^L_0(p^2))(1+\Pi^R_0(p^2))}|\mathcal{F}_t(h,\eta)|^2\right],
    \label{eq:Vt1}
\end{align}
where $N_c=3$ is the number of colors. The expression for $\mathcal{F}_t(h,\eta)$ in unitary gauge is given by
\begin{align}
    \mathcal{F}_t(h,\eta) = \frac{h s_{\pi_0}}{\pi_0^2}\left[c_\beta \left(\pi_0 c_{\pi_0} - i \eta s_{\pi_0}\right) - e^{i\gamma} s_\beta \left(\pi_0 c_{\pi_0} + i \eta s_{\pi_0}\right)\right],
    \label{eq:f(h,eta)}
\end{align}
where $\pi_0\equiv \sqrt{h^2+\eta^2}$, $c_{\pi_0} (s_{\pi_0}) \equiv \cos\pi_0/f (\sin\pi_0/f)$, and $c_\beta (s_\beta) \equiv \cos\beta (\sin\beta)$. The angle $\beta$ and the phase $\gamma$ arise from the projection matrices $A_{L,R}$ used for embedding the elementary quarks in $\mathbf{6}$ of $SU(4)$ (see \cref{model_details}). 


We perform the integration over Euclidean momentum in \cref{eq:Vt1}, retaining only the leading term in the logarithmic expansion. In section \ref{pot_thermal}, we derive the full analytic expression for the potential and discuss the validity of the leading logarithm (LL) approximation. From the definitions of $\Pi^{L,R}_0$ and $\Pi^{LR}_2$ given in \cref{model_details}, it is clear that the integration in \cref{eq:Vt1} is UV finite. This finiteness is not coincidental but is attributed to the presence of maximal symmetry. Without this symmetry, additional contributions to the potential would emerge, whose coefficients involve the following quadratically divergent integral in the LL approximation:
\begin{align}
    \int \frac{d^4p}{(2\pi)^4}\frac{\Pi^{L,R}_1}{1+\Pi^{L,R}_0}\,, \quad {\rm where} \quad \lim_{p^2\to \infty} \frac{\Pi^{L,R}_1}{1+\Pi^{L,R}_0} \sim \frac{1}{p^2}\,. 
\end{align}
However, due to the maximal symmetry, such potentially dangerous integrals vanish identically since the form factor $\Pi^{L,R}_1$ becomes zero owing to the relation in \cref{eq:maxsym}, see \cref{model_details} for the expression of $\Pi^{L,R}_1$.

Using the form factors defined in \cref{form__factors}, the LL convergent expression for the top-induced zero temperature potential is 
\begin{align}
    V_t \approx -2N_c \int \frac{d^4p}{(2\pi)^4} \frac{|F_t|^2 |\mathcal{F}_t(h,\eta)|^2 M_L^2 M_R^2}{p^2(p^2+M_L^2)(p^2+M_R^2)} = -\frac{N_c}{8\pi^2}|F_t|^2 M_L M_R K\left(\frac{M_L}{M_R}\right) |\mathcal{F}_t(h,\eta)|^2,
    \label{top_pot}
\end{align}
where
\begin{align}
    F_t=\frac{\lambda^L\lambda^{R*} f^2 M_5}{M_L M_R}, \quad M_{L,R}=\sqrt{M_5^2+|\lambda^{L,R}|^2f^2}, \quad {\rm and} \quad K(x)\equiv \frac{2x}{x^2-1}\ln x.
    \label{kx}
\end{align}

\subsubsection*{Gauge contribution}
\label{pot_zero_gauge}

Proceeding in a similar manner, the gauge contribution $V_g$ is calculated using \eqref{gauge_lag} as 
\begin{align}
    V_g = \frac{6}{2} \int \frac{d^4p}{(2\pi)^4} \log\left[1 + \frac{\Pi_1(p^2)}{\Pi^W_0(p^2)} \mathcal{F}_g(h,\eta) \right] + \frac{3}{2} \int \frac{d^4p}{(2\pi)^4} \log\left[1 + \left(\frac{\Pi_1(p^2)}{\Pi^W_0(p^2)}+\frac{\Pi_1(p^2)}{\Pi^B_0(p^2)}\right) \mathcal{F}_g(h,\eta) \right].
    \label{eq:V_g0_ex}
\end{align}
The first integral arises from the $W^{\pm}_\mu$ bosons and the second from the $Z_\mu$ boson. The function $\mathcal{F}_g(h,\eta)$ is defined as
\begin{align}
    \mathcal{F}_g(h,\eta) = \frac{h^2}{4\pi_0^2}s_{\pi_0}^2\,.
\end{align}

The leading momentum dependence of the form factors $\Pi^{W,B}_0$, as shown in \cref{approximation_1} is given by $\Pi^{W,B}_0\sim p^2$. Thus in the LL approximation, the integrals over the form factors in \cref{eq:V_g0_ex} are UV convergent if $\Pi_1 \sim 1/p^4$ at large $p^2$. To ensure the large $p^2$ behavior of $\Pi_1(p^2)$ as shown in \cref{eq:Pi_1_g_form}, we impose two sum rules \cref{sumrules}, resulting in the final expression presented in  \cref{eq:Pi_1_g_WSR} of the \cref{model_details}. Employing \cref{eq:Pi_1_g_WSR}, the integrated expression for the gauge contribution at the LL approximation is

 \begin{align}
     V_g \approx \frac{3}{2} (3 g^2+g^{\prime 2}) \int \frac{d^4p}{(2\pi)^4} \frac{f^2 \mathcal{F}_g(h,\eta) m_\rho^2 m_a^2}{p^2(p^2+m_\rho^2)(p^2+m_a^2)} = \frac{3}{32\pi^2} (3 g^2+g^{\prime 2}) f^2  m_{\rho}m_a  K\left(\frac{m_\rho}{m_a}\right) \mathcal{F}_g(h,\eta)\,.
     \label{gauge_pot}
 \end{align}

\subsubsection*{Hyperquark contribution}
\label{pot_zero_hyperquark}

We choose the hyperquark mass term such that it breaks $SU(4)$ explicitly, while leaving $SU(2)_L\times SU(2)_R$ unbroken
\begin{align}
    \mu_H = i\, m\left(\begin{array}{cccc}
        \sigma^2 & 0\\
        0 & -e^{i\delta} \sigma^2
    \end{array}\right). 
\end{align}
Here, $m$ is taken to be real and $\delta$ denotes a generic complex phase signalling that CP is broken in the strong sector. The hyperquark mass contribution to the pNGB potential is given by \cite{Dong:2020eqy,Frandsen:2023vhu}
\begin{align}
    V_H = B f^3 {\rm tr} \left[ \mu_H U + U^\dagger \mu_H^\dagger \right] = F_H \left[(1+c_\delta)c_{\pi_0} + s_\delta \frac{\eta}{\pi_0}s_{\pi_0}\right]\,, 
\end{align}
where $c_\delta (s_\delta) \equiv \cos\delta (\sin\delta)$, $B$ is a dimensionless parameter, and $F_H\equiv 4B m f^3$. Note that the hyperquark contribution preserves maximal symmetry, since $\mu_H$ transforms as $\mu_H\to g_L^*\mu_H g_R^\dagger$.

\subsection*{Summary of zero temperature effective potential}
\label{summary_V0}
Here we summarize the entire zero temperature effective potential at one loop for the convenience of the readers. In the LL approximation, the potential $V^{LL}_{1\text{-loop}}$ is given by three separate contributions as 
\begin{equation}
    V^{LL}_{1\text{-loop}} =  - F_T |\mathcal{F}_t(h,\eta) |^2 + F_G \mathcal{F}_g(h,\eta) + F_H \mathcal{F}_H(h,\eta). \label{LL_potential}
\end{equation}
The field dependent functions $\mathcal{F}_i(h,\eta)$ are defined as 
\begin{align}
\nonumber
\mathcal{F}_t(h,\eta)  & \equiv \frac{h s_{\pi_0}}{\pi_0^2}\left[c_\beta \left(\pi_0 c_{\pi_0} - i \eta s_{\pi_0}\right) - e^{i\gamma} s_\beta \left(\pi_0 c_{\pi_0} + i \eta s_{\pi_0}\right)\right], \\
\mathcal{F}_g(h,\eta)  & \equiv \frac{h^2}{4\pi_0^2}s_{\pi_0}^2, \qquad
\mathcal{F}_H(h,\eta)  \equiv \left[(1+c_\delta)c_{\pi_0} + s_\delta \frac{\eta}{\pi_0}s_{\pi_0}\right],
\end{align}
and the prefactors are
\begin{align}
F_T\equiv\frac{N_c}{8\pi^2}|F_t|^2 M_L M_R K\left(\frac{M_L}{M_R}\right), \quad
F_G\equiv \frac{3}{32\pi^2} (3 g^2+g^{\prime 2}) f^2  m_{\rho}m_a  K\left(\frac{m_\rho}{m_a}\right),\quad
F_H\equiv4 B m f^3. 
\label{F_coefficients}
\end{align}

The total LL pNGB potential \eqref{LL_potential} is fully characterized by the coefficients of each source of explicit symmetry breaking, which we have denoted by $F_T, F_G, F_H$, corresponding to the top, gauge and hyperquark contributions, respectively. Additionally the angle $\beta$ in the embedding of the top quark, and the CP violating phases $\gamma$, $\delta$ from the top and the hyperquark contributions also appear, while the trigonometric functions of the fields are modulated by the pNGB decay constant $f$. Thus a total of seven free parameters need to be specified to investigate the vacuum structure of the model where the singlet obtains a non-zero vev.

\subsection{Masses and mixing}
\label{spectra}

The Lagrangian of this model involves two complex phases, namely $\gamma$ and $\delta$, which lead to CP violation. In the absence of these phases, the pNGB potential exhibits a $\mathbb{Z}_2$ symmetry under which $h\to h$ and $\eta \to -\eta$. It has been argued in \cite{Alanne:2018wtp} that this $\mathbb{Z}_2$ symmetry cannot be broken spontaneously by giving a vev to $\eta$ unless an explicitly  CP violating phase is present in strong sector. The global minimum of the potential is determined by the relative size of explicit $SU(4)$ breaking coefficients (captured by the parameters $F_T$, $F_G$ and $F_H$) and the angles and phases ($\beta$, $\gamma$, and $\delta$) appearing inside the $\mathcal{F}(h,\eta)$ functions in \cref{LL_potential}. We are interested in a zero temperature vacuum where both $h$ and $\eta$ receive vevs. The presence of a tadpole term for $\eta$ in $V_H$ (through the function $\mathcal{F}_H$) proportional to $\sin\delta$ leads to a non-zero $v_\eta$, which is physical due to the explicit CP violation in the strong sector as well as in the top sector in the presence of complex phases $\delta$ and $\gamma$, respectively. Although, $V_H$ plays a crucial role in giving a non-zero $v_\eta$, the numerical value of $|F_H|$ is required to be small compared to $F_T$ and $F_H$ to reproduce the correct EW vacuum, as will be shown in \cref{FOPT}.

The masses of the weak gauge bosons and top quark  are obtained by taking the limit $p^2\to 0$ in \eqref{gauge_lag} and \eqref{top_lag}, respectively, and replacing $(h,\eta)$ with their respective vacuum expectation values $(v,v_\eta)$ as
\begin{align}
   M_W^2=M_Z^2 c_W^2 = g^2\Pi_1(0)\mathcal{F}_g(v,v_\eta) =\frac{g^2v^2_{\rm EW}}{4}\,, \quad  m_t = \left|\frac{\Pi^{LR}_2(0)\mathcal{F}_t(v,v_\eta)}{\sqrt{(1+ \Pi^L_0(0)) (1+ \Pi^R_0(0)) }}\right| = F_t |\mathcal{F}_t(v,v_\eta)|,
    \label{mwmt}
\end{align}
where $c_W^2\equiv \cos\theta_W^2$ is the cosine of the Weinberg angle and the EW vev is defined as
\begin{align}
    \quad v_{\rm EW} = \frac{f v}{\sqrt{v^2+v_\eta^2}}\sin\frac{\sqrt{v^2+v_\eta^2}}{f} = 246~ {\rm GeV}\,. \label{h_vev}
\end{align}
In presence of a non-zero $v_\eta\ne 0$, a mixing between CP even $h$ and CP odd $\eta$ arises due to off-diagonal kinetic and mass terms. Canonical normalization of the kinetic terms of $h$ and $\eta$ is discussed in \cref{kinetic_mixing}, while the mass diagonalization leads to a further mixing between them by an angle $\theta$. The relation between the scalars in the mass basis $(\hat{h},\hat{\eta})$ with the gauge basis is given by 
\begin{equation}
    \begin{pmatrix}
    h\\
    \eta \\
    \end{pmatrix} 
    =
    \begin{pmatrix}
    a & 0\\
    b & c\\
    \end{pmatrix}
    \begin{pmatrix}
    \cos\theta & -\sin\theta\\
    \sin\theta & \cos\theta\\
    \end{pmatrix}
    \begin{pmatrix}
    \hat{h}\\
    \hat{\eta}\\
    \end{pmatrix}\,, 
\end{equation}
where $a,b,c$, required for canonical normalization of the kinetic terms, are defined in \cref{kinetic_mixing}. The overall scale of the potential can be fixed by identifying the mostly doublet like state $\hat{h}$ with the observed 125 GeV Higgs boson.

\section{pNGB potential at finite temperature}
\label{pot_thermal}

The computation of the thermal potential in maximally symmetric composite Higgs model has a crucial difference from the conventional calculations involving elementary scalar fields. The zero temperature potential of the pNGBs due to contributions from the top quark and gauge bosons arises at the one-loop level, except for the hyperquark mass contribution. Thus, unlike the tree-level elementary scalar potential, both the zero temperature and thermal potentials emerge at the same loop order. In what follows, we perform an exact analytic computation of the full one-loop contributions to the pNGB potential from the top and gauge sector at finite temperature.

We incorporate the effects of momentum-dependent form factors in the computation of the finite temperature potential to show that the contributions from the heavy top-partners and the spin-1 resonances are exponentially suppressed at finite temperature. As a result, we retrieve the standard results for the thermal potential \cite{Dolan:1973qd}, in terms of the field dependent mass of the top quark and the weak gauge bosons.

As a bonus, in the $T=0$ limit our computation of the thermal potential yields an exact analytic expression for the zero temperature potential originating from the gauge and top contributions, which to the leading order matches with the results obtained in \cref{pot_zero}. This innovative approach to derive an exact analytic form of both the zero temperature and finite temperature potential can be readily extended to other composite Higgs models, although the final result may not be convergent if maximal symmetry is not imposed.

The generic structure of the one-loop potential from the top quark (\cref{top_pot}) and the gauge (\cref{gauge_pot}) sector can be written as
\begin{equation}
V_{1\text{-loop}}=
 \frac{N_{\rm eff}}{2}\int \frac{d^4p}{(2\pi)^4} \log\left[1 + \frac{m^2_{\text{SM}} m^2_1 m^2_2}{p^2(p^2+m_1^2)(p^2+m_2^2)} \right],
 \label{eq:V^T_gen_1}
\end{equation}
where $N_{\rm eff}$, and $m_{\rm SM}$ denote the number of effective degrees of freedom and the field dependent mass of the SM particles (top quark or W,Z bosons), respectively, while $m_{1,2}$ are the resonance masses. In particular, $(N_{\rm eff},m_{\rm SM}^2,m_{1,2})$ are given by $(-N_c,|F_t|^2|\mathcal{F}_t(h,\eta)|^2,M_{L,R})$  for the top quark contributions, $(6,g^2f^2\mathcal{F}_g(h,\eta),m_{\rho,a})$ for the W-boson loop, and $(3,g^2c_W^2f^2\mathcal{F}_g(h,\eta),m_{\rho,a})$ for the Z-boson loop. Completing the sum inside the logarithm, we arrive at a more convenient form of the potential where the numerator in the argument of the log is given by a cubic polynomial in $p^2$ as
\begin{align}
V_{1\text{-loop}} =\frac{N_{\rm eff}}{2}\int \frac{d^4p}{(2\pi)^4} \log\left[\frac{p^2(p^2 + m^2_1)(p^2 + m^2_2) + m^2_{\text{SM}} m^2_1 m^2_2}{p^2(p^2+ m^2_1)(p^2+m^2_2)}\right].
 \label{eq:V_gen_inter_2}
\end{align}
The numerator inside the argument of the logarithm can be factorized to give
\begin{align}
V_{1\text{-loop}}= \frac{N_{\rm eff} }{2}\int \frac{d^4p}{(2\pi)^4} \log\left[\frac{(p^2+\tilde{m}_1^2)(p^2+\tilde{m}^2_2)(p^2+\tilde{m}^2_3)}{p^2(p^2+ m^2_1)(p^2+m^2_2)}\right],
    \label{eq:V_genI}
\end{align}
where $\tilde{m}_i^2$ ($i=1,2,3$) are obtained by comparing the coefficients of different powers of $p^2$ in the numerators of \cref{eq:V_gen_inter_2} and \cref{eq:V_genI}.
They satisfy the  Vi\`ete equations, given by
\begin{align}
    \tilde{m}^2_1 \tilde{m}^2_2 \tilde{m}^2_3=m^2_{\text{SM}} m_1^2 m_2^2,\quad  
    \tilde{m}^2_1 +\tilde{m}^2_2 +\tilde{m}^2_3= m_1^2+ m_2^2, \quad 
    \tilde{m}^2_1 \tilde{m}^2_2 +\tilde{m}^2_1\tilde{m}^2_3+\tilde{m}^2_2\tilde{m}^2_3= m_1^2 m_{2}^2\,.
    \label{eq:ViettIII}
\end{align}
It follows that $\tilde{m}_1^2=m^2_{\text{SM}}$, $\tilde{m}_2^2=m^2_{1}$ and $\tilde{m}_3^2=m^2_{2}$ approximately solve the above equations, up to $\mathcal{O}(m^2_{\text{SM}}/m^2_{1,2})$ terms. Note that each solution to Vi\'ete equations \eqref{eq:ViettIII} must be either real or complex conjugate of some other solution. In the limit $m^2_{\text{SM}}\ll m^2_{a,\rho}$, the real parts of $\tilde{m}^2_i$ are different, implying that there is no conjugate pair among the solutions and they are all real-valued.

Thermal corrections to the pNGB potential are computed using the imaginary time formalism which amounts to the following substitution in the momentum integrals
\begin{equation}
\int dp^0 d^3 p \ f(p^2) \rightarrow 2\pi T \sum_{n=-\infty}^{\infty} \int d^3 p\ f(\omega_n^2+|\vec{p}|^2)\,, 
\label{Imaginary_formalism}
\end{equation}
where the $n^{\rm th}$ Matsubara frequency is given by $\omega_n = 2\pi n T$ ($\omega_n = (2 n+1)\pi  T$) for bosons (fermions), and $f(p^2)$ is some momentum-dependent function (see  \cite{Sher:1988mj,Quiros:1999jp,Hindmarsh:2020hop} for comprehensive reviews). For convenience, we focus on the bosonic integral, while the same procedure can be followed for the case of fermionic contributions just by replacing the appropriate Matsubara frequencies. Using \cref{Imaginary_formalism} in \cref{eq:V_genI} and replacing $m_{1,2}\to m_{\rho,a}$ we get
\begin{align}
V_{1\text{-loop}}= \frac{N_{\rm eff} }{2}T\int \frac{d^3p}{(2\pi)^3} \log\left[\prod_{n\in\mathbb{Z}}\frac{(\alpha^2\tilde{E}_1^2+n^2)(\alpha^2\tilde{E}_2^2+n^2)(\alpha^2\tilde{E}_3^2+n^2)}{(\alpha^2|\vec{p}|^2+n^2)(\alpha^2 E^2_\rho+ n^2)(\alpha^2 E^2_a+n^2)} \right],
    \label{eq:V^T_genI}
\end{align}
where $\tilde{E}^2_{i}\equiv|\vec{p}|^2 + \tilde{m}^2_{i}$, $E^2_{\rho,\; a}\equiv|\vec{p}|^2 + m^2_{\rho,\; a}$, and $\alpha\equiv 1/(2\pi T)$. 
Exploiting the convergence of the integral in \eqref{eq:V^T_gen_1} we interchange the order of integration and the summation over $n$ in the above equation, and further express the infinite sum of the logarithms as logarithm of product of its arguments.


The product in \cref{eq:V^T_genI} can be directly evaluated as\footnote{One can express $\prod_{n\in\mathbb{Z}}(n^2+x^2)/(n^2+y^2)=\prod_{n\in\mathbb{Z}}(n^2+x^2)/n^2\times\prod_{n\in\mathbb{Z}}n^2/(n^2+y^2)$, which is equivalent to the product expansion of $\sinh^2(\pi x)/\sinh^2(\pi y)$ for any $x$ and $y$ in the complex plane.}. 
\begin{align}
V_{1\text{-loop}}&=    \frac{N_{\rm eff}}{2}T\int \frac{d^3p}{(2\pi)^3} \log\left[   \frac{ \prod_{i=1}^{3} \sinh^2{(\pi \alpha\tilde{E}_i)}  }{ \sinh^2{(\pi \alpha |\vec{p}|)} \sinh^2{(\pi \alpha E_{\rho})} \sinh^2{(\pi \alpha E_{a})} }  \right] \nonumber \\
& =N_{\rm eff} \sum^3_{i=1}\int \frac{d^3p}{(2\pi)^3}\left[\frac{\tilde{E}_i}{2}+T\log\left(1-e^{\tilde{E}_i/T}\right)\right] + (\text{field-independent terms}), 
    \label{eq:V^T_genII}
\end{align}
where in the second line, we once again exchange the summation and the integration, and separate the $T=0$ part from the finite temperature corrections. The final closed form expression for the one-loop potential at finite temperature (up to constant, field-independent terms) is given by  
\begin{equation}
 V_{1\text{-loop}} = 
 V^{(T=0)}_{\text{CW}}(\tilde{m}_i)+N_{\rm eff}\frac{T^4}{2\pi^2} \sum^3_{i=1} J_B\left(\frac{\tilde{m}_i}{T} \right),
\label{eq:V^T_genIV}
\end{equation}
where
\begin{equation}
 J_B(x) \equiv \int_0^{\infty}dy\, y^2 \log\left[1-e^{-\sqrt{y^2+x^2}}\right],
\label{eq:JB}
\end{equation}
and the zero temperature potential is
\begin{align}
V^{(T=0)}_{\text{CW}}(\tilde{m}_i)  \equiv \frac{N_\text{eff}}{2}\int \frac{d^3p}{(2\pi)^3}\sum^3_{i=1}\tilde{E}_i = \frac{N_{\rm eff}}{32\pi^2}\sum_{i=1}^3\tilde{m}^4_i\log\left(\frac{\tilde{m}_i}{\mu}\right) 
 . \label{eq:V_ex}
\end{align}
We introduce an arbitrary scale $\mu$ to keep the arguments of the logarithms dimensionless. Nevertheless we emphasize that the potential is completely independent of the choice of $\mu$ as we show in \cref{app:Viett}. Remarkably, \cref{eq:V^T_genIV,eq:V_ex} yield the familiar expressions for the one-loop thermal potential, derived in \cite{Dolan:1973qd}. The crucial difference is in $V^{(T=0)}_{\text{CW}}$, which unlike an elementary scalar theory, is convergent for maximally symmetric composite Higgs model.

Recall that the field dependence in the one-loop potential is encoded inside the quantities $\tilde{m}_i$. In the limit $m_{\rho,a}^2 \gg m_{\rm SM}^2$, two ``heavy'' solutions of  \cref{eq:ViettIII}, $\tilde{m}_i^2\simeq m_{\rho,a}^2(1 + \mathcal{O}(m_{\rm SM}^2/m^2_{\rho,a}))$ contribute to the field dependent terms of $V^{(T=0)}_{\text{CW}}$ through the next-to-leading order $\mathcal{O}(m^2_{\text{SM}}/m^2_{\rho,\;a})$ terms as $m^2_{\rho,\;a}m^2_{\text{SM}}$. In comparison, the ``light'' solution of \cref{eq:ViettIII}, $\tilde{m}_i^2\simeq m_{\rm SM}^2$, contributes to $V^{(T=0)}_{\text{CW}}$ as $m_{\rm SM}^4 \ll m^2_{\rho,\;a}m^2_{\text{SM}}$. Thus, the zero temperature potential is dominated by the next-to-leading order terms of the ``heavy'' solutions of \cref{eq:ViettIII}, which are precisely captured by the LL expansion introduced in \cref{pot_zero}. We have verified that, for $m_{\rho,a} > f$, the  difference between the exact solution and the LL expansion is sufficiently small (below $5\%$ in all relevant cases), prompting us to use the latter for the numerical scan of the parameter space. This makes our large-scale numerical scans much faster since the full expression for the potential requires solving the equations \eqref{eq:ViettIII} for each set of benchmark parameters and across every field space coordinate in which the potential needs to be evaluated.

In contrast to the $T=0$ case, the finite temperature part of the scalar potential is dominated by the ``light'' solution $\tilde{m}^2_i\sim m_{\text{SM}}^2$ of \cref{eq:ViettIII}. The impact of the ``heavy'' solutions $\tilde{m}_i\simeq m_{\rho,a}$ on the thermal potential are, instead, suppressed by $\exp(-m_{\rho,a}/T)$. Hence, the finite temperature corrections from the bosonic loops simply reduce to the usual expressions for an elementary scalar theory:
\begin{equation}
    V_{\text{Bosons}}^{T}(h,\eta)\approx N_{\text{eff}}\frac{T^4}{2\pi^2} J_B\left(\frac{{M}_{W,Z}}{T}\right),
    \label{bosonic_thermal}
\end{equation}
where we restore $m_{\rm SM}=M_{W,Z}(h,\eta)$. A similar computation can be performed to evaluate the fermionic (top quark) contributions, yielding analogous result
\begin{equation}
    V_{\text{Fermions}}^{T}(h,\eta)\approx N_{\text{eff}}\frac{T^4}{2\pi^2} J_F\left(\frac{{m}_{t}}{T}\right), \quad {\rm where},\quad J_F(x)\equiv -\int_0^{\infty}dy\, y^2 \log\left[1+e^{-\sqrt{y^2+x^2}}\right]. \label{fermionic_thermal}
\end{equation}

\subsection*{Further comments on the finite temperature potential}

Before ending this section, we highlight some crucial aspects of the thermal potential calculated above. 

\begin{itemize}

\item The exact analytic expressions for the one-loop potential at finite temperature including the non-trivial momentum-dependent form factors yield the same functional form, as in \cite{Dolan:1973qd}, with the standard thermal functions, (see \cref{bosonic_thermal,fermionic_thermal}). Contributions from the heavy resonances are exponentially suppressed in the limit $m_{a,\rho}\gg T$, which is aligned with our scenario where the nucleation temperature for the phase transition is approximately $T_n \approx \mathcal{O}(100)$ GeV, while $m_{a,\rho}\gtrsim 1$ TeV. Thus, the leading corrections to the potential at finite temperature primarily depend on the field dependent masses of W,Z bosons and top quark, while the resonance masses play an important role to set the masses and vevs of the SM fields at zero temperature. 

\item Additionally, we have shown that the imaginary time formalism can be used to evaluate the thermal potential even when  complicated momentum dependence appears inside the one-loop integral. This novel calculation can in principle be applied to other composite Higgs models, as we show in the \cref{app:V_gen}. At zero temperature the  LL result,  \cref{LL_potential}, gives a fairly accurate approximation of the exact potential. Moreover, we stress that the potential at $T=0$ is finite owing to the maximal symmetry, which is in contrast to a theory involving elementary scalar fields in which infinities are absorbed into renormalization constants.

\item Notably, the scalar self-interactions of the pNGBs arise at one loop, which implies that their contribution to the thermal potential is a two-loop effect. However, if the zero temperature masses of the scalars are of the same order as the temperature of the PT, the two-loop effects might be  sizeable. A full computation of the two-loop thermal potential is beyond the scope of this study, nevertheless, we anticipate that the contributions from the self-interactions of the pNGBs are comparatively small, because the number of pNGB degrees of freedom in unitary gauge, which contribute to the thermal potential, is significantly less than those of the top quark and weak gauge bosons.

\item  Thermal effects from the hyperquark condensate can lead to corrections to the effective potential. In the chiral limit ($\mu_H\to 0$)~\footnote{In section \ref{FOPT} we show that for viable parameter values, the Hyperquark mass $m\sim F_H/f^3$ is far below the transition temperature, so the chiral approximation is valid.} the scale of symmetry breaking in the strong sector receives a correction as $f(T)\sim f(1+\mathcal{O}(T^2/f^2))$~\cite{Gasser:1986vb}. In our model, this contribution is numerically suppressed at the temperature of the PT which is at the electroweak scale, much below $f\sim$ TeV, and therefore we neglect this correction. 

\item The perturbative computation of the finite temperature potential suffers from well-known IR divergences, which can be mitigated by resumming the most divergent classes of diagrams. In theories with fundamental scalars this can be achieved by incorporating effective daisy terms to the tree-level bosonic masses that contribute to the one-loop potential. However, a consistent resummation in composite Higgs model is a complex endeavor warranting a separate study. Instead, we use the SM expressions for the daisy terms which can be found for instance in~\cite{Beniwal:2017eik}. 

\end{itemize}

\section{Phase transition and gravitational waves}
\label{FOPT}

In this section we discuss the thermal evolution of the pNGB potential and the possibility of a FOPT. We further demonstrate that stochastic GW signatures may result from such strong FOPT, which can in principle be probed by LISA and AEDGE in the future. 

\subsection{Experimental constraints on model parameters}
\label{LHC}

In order to conduct a numerical analysis of the zero temperature potential and its thermal evolution, it is necessary to fix a suitable range for the free parameters. Constraints from the LEP, LHC and EDM measurements are taken into account to determine the allowed range of parameters. 

At low energy (in the $p^2\to 0$ limit), interactions of the pNGBs with the weak gauge bosons and the top quark, in the mass basis, are given by
\begin{align}
    \mathcal{L}= \left(g_{\hat{h}}\hat{h} + g_{\hat{\eta}}\hat{\eta} \right)\left[W^+_\mu W^{-\mu}+\frac{1}{2c_w^2}Z_\mu Z^\mu\right] + \hat{h}\Bar{t}\left[{\rm Re}(y_{\hat{h}})+i {\rm Im}(y_{\hat{h}})\gamma_5\right]t + \hat{\eta}\Bar{t}\left[{\rm Re}(y_{\hat{\eta}})+i {\rm Im}(y_{\hat{\eta}})\gamma_5\right]t,
     \label{CP_violation}
\end{align}
where the coupling strengths are calculated as
\begin{align}
   g_{\hat{h}} & = g^2 f^2\mathcal{D}_h \mathcal{F}_g(v,v_\eta)\,, & y_{\hat{h}} & = F_t\mathcal{D}_h\mathcal{F}_t(v,v_\eta)\,,  & \mathcal{D}_h & \equiv \left[a c_\theta \partial_h +\left(b c_\theta + c s_\theta\right) \partial_\eta \right],
   \\
    g_{\hat{\eta}} & = g^2 f^2\mathcal{D}_\eta \mathcal{F}_g(v,v_\eta)\,, & 
    y_{\hat{\eta}} & = F_t\mathcal{D}_\eta\mathcal{F}_t(v,v_\eta)\,, & \mathcal{D}_\eta & \equiv \left[-a s_\theta \partial_h +\left(-b s_\theta + c c_\theta\right) \partial_\eta \right].
\end{align}

The presence of CP violating couplings of $\hat{h}$ with the top quark is a direct consequence of CP violation in the strong sector and mixing between the CP even $h$ and CP odd state $\eta$. Below we summarize the experimental limits considered for the rest of this analysis:

\begin{itemize}
    \item We demand that at the zero temperature the potential is minimized at the EW vacuum, reproducing the observed values for the masses of the Higgs boson, the top quark and the weak gauge bosons. Using \cref{mwmt}, the top quark mass is fixed within a range between $[140,175]$ GeV. We further ensure that the eigenstate with the mass of 125 GeV is mostly $SU(2)_L$ doublet-like, so that it can be identified with the observed Higgs boson. For all the viable parameter space that satisfies the above requirements we observe that $F_G \gg F_T$ holds. We will see that requiring FOPTs yield $F_G \gg F_T \gg |F_H|$. 
    
    \item EW precision data \cite{Contino:2010rs,Agashe:2005dk,Falkowski:2013dza,Contino:2015mha} provides a strong bound on the decay constant of the pNGBs $f\gtrsim 1$ TeV, however its precise value depends on the specific UV completion and can be somewhat relaxed. On the other hand, current measurements at the LHC allow for around 5\% deviation of the $hWW$ and $hZZ$ couplings and around 10\% deviation in the $ht\Bar{t}$ couplings \cite{ATLAS:2022vkf,CMS:2022dwd}. These limits can be translated into a bound on $f$ which is also in the same ballpark range of a TeV. Considering both the EW precision measurements and the Higgs coupling measurements at the LHC, we choose $f=1$ TeV. Note that the doublet-singlet mixing between $h$ and $\eta$ leads to further modifications of the Higgs couplings in addition to those appearing from the pNGB nature of the Higgs. To tame this additional modification, we restrict the mixing angle within $|\sin\theta|\lesssim 0.2$.

        \begin{figure}[t]
    \centering
    \includegraphics[width=0.468\textwidth, keepaspectratio]{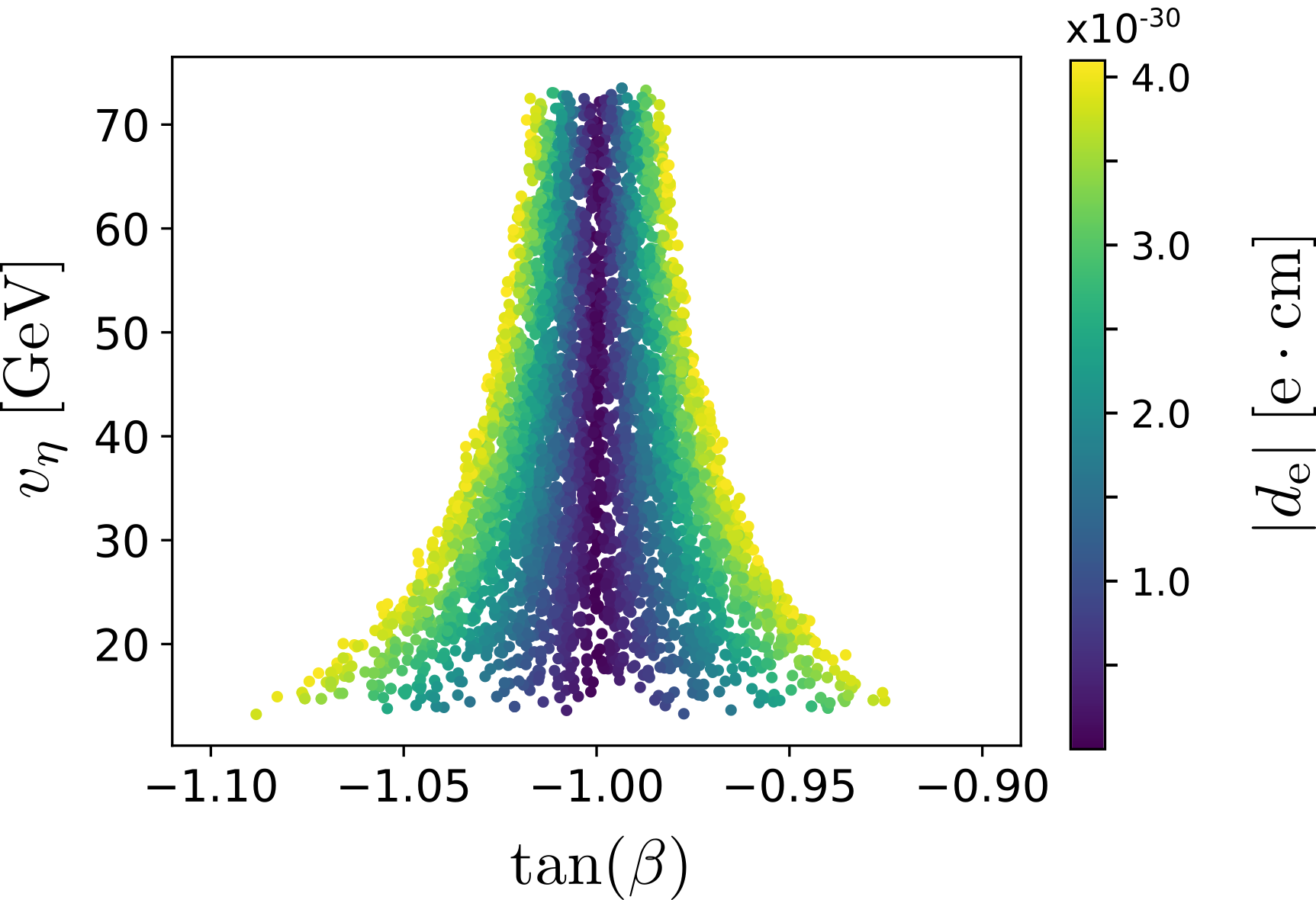}\hspace{5mm}
    \includegraphics[width=0.492 \textwidth, keepaspectratio]{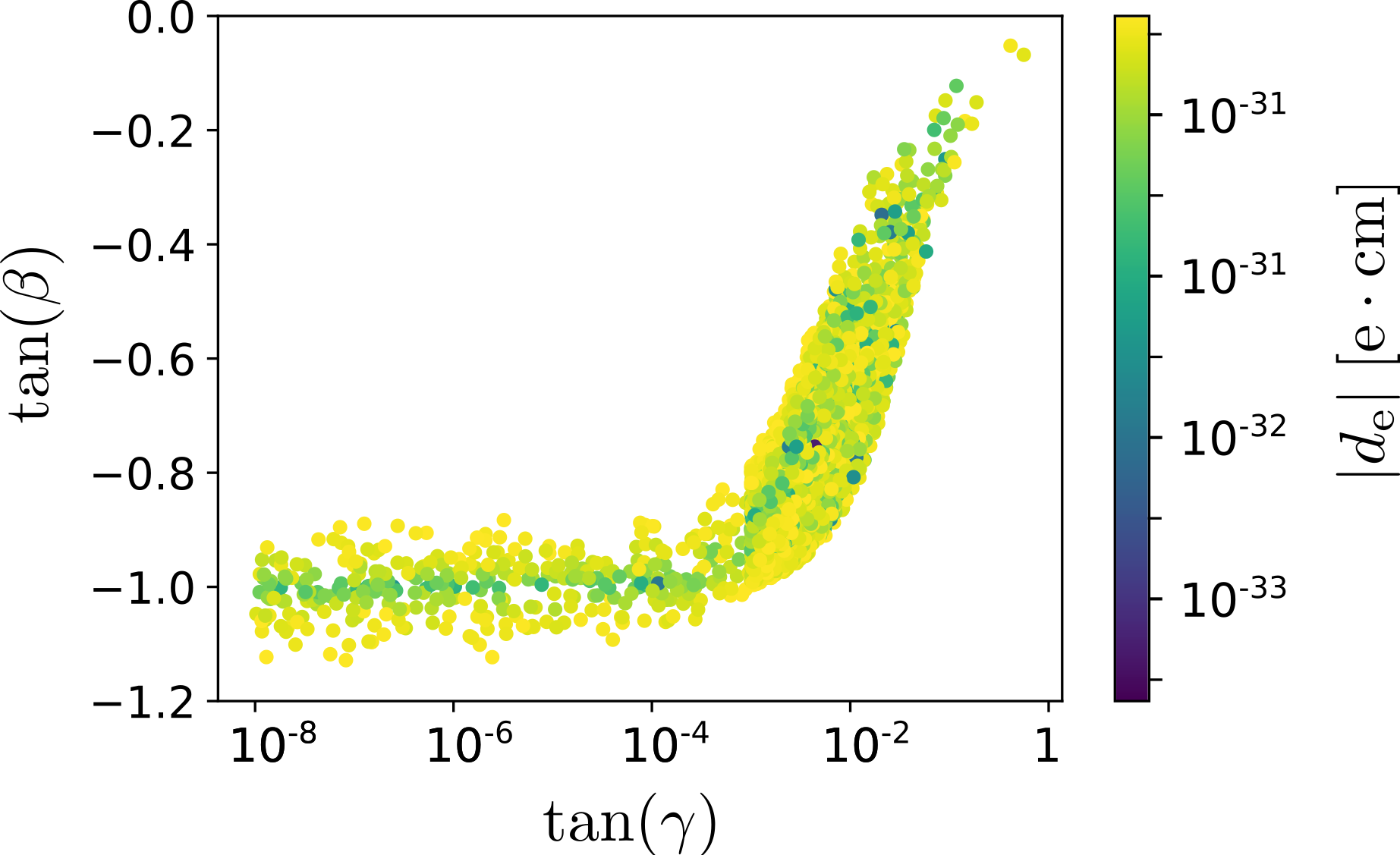}
    \caption{\sf\it Benchmarks satisfying the upper limit on the electron EDM, $d_e<4.1\times10^{-30} \, e\, \text{cm}$~\cite{Roussy:2022cmp} in the $\tan(\beta)-v_\eta$ plane for $\gamma=0$ (left panel), and in the $\tan(\beta)-\tan(\gamma)$ plane (right panel).}
    \label{fig:tan(beta)-vs}
\end{figure}

    \item Non-observation of any new particles at the LHC leads to strong lower bounds on the masses of the top-partners, see \cite{Banerjee:2022xmu,Banerjee:2024zvg} for a review of the current bounds. Similarly, the masses of the spin-1 resonances are also constrained from both direct searches at the LHC as well as from the  EW precision measurements \cite{Ghosh:2015wiz, Aaboud:2017cxo, Low:2015uha,Niehoff:2015iaa,Franzosi:2016aoo,Liu:2018hum, Banerjee:2021efl}. However, it is important to note that the experimental bounds are provided under the assumption that these resonances can only decay into SM 2-body final states. For the model under consideration, due to the presence of a pNGB singlet $\eta$, new decay channels of the resonances may open up \cite{Benbrik:2019zdp}, leading to a  relaxation of the current bounds. Considering these possibilities, we take $M_5\geq 1$ TeV and $m_{\rho,a}\geq 1.5$ TeV. To circumvent the limit from the invisible branching ratio of the Higgs boson, we forbid the $h\to \eta\eta$ decay channel by considering $m_\eta \geq 2m_h$.

    \item The presence of explicitly CP violating phases $\gamma, \delta$ and a non-zero $v_\eta$ of the CP odd scalar $\eta$ lead to non-zero electron and neutron EDMs through two-loop Barr-Zee diagrams \cite{Barr:1990vd,Keus:2017ioh}. At present, the strongest constraint on the CP violating phases comes from the measurement of the electron EDM. The experimental upper bounds on the electron EDM is given by the ACME collaboration as $|d_e|<1.1 \times 10^{-29} e \ \text{cm}$ \cite{ACME:2018yjb}, while a more recent and independent measurement set the bound to $|d_e|<4.1\times10^{-30}e \ \text{cm}$ \cite{Roussy:2022cmp}. In this paper we only present results consistent with the latter bound. We divide the parameter space in two regions. First we fix $\gamma\simeq 0$ to highlight the impact of the CP violation arising solely from the phase $\delta$ in the hyperquark mass. In this case, the angle $\beta$ is strongly constrained by the EDM bound to be around $\beta\sim 3\pi/4$, as shown in the left panel of \cref{fig:tan(beta)-vs}. Next, to assess the impact of $\gamma$, we show the allowed range of $\tan(\beta)$ and $\gamma$ satisfying the EDM constraint on the right panel of \cref{fig:tan(beta)-vs}.

    The CP violating coupling of the Higgs boson with the top quark is also constrained by the LHC data, however those limits are comparatively weaker than the limit from the electron EDM. 
    \end{itemize}

\subsection*{Scanning parameter space}

\begin{figure}[t]
    \centering
    \includegraphics[width=0.65\textwidth, keepaspectratio]{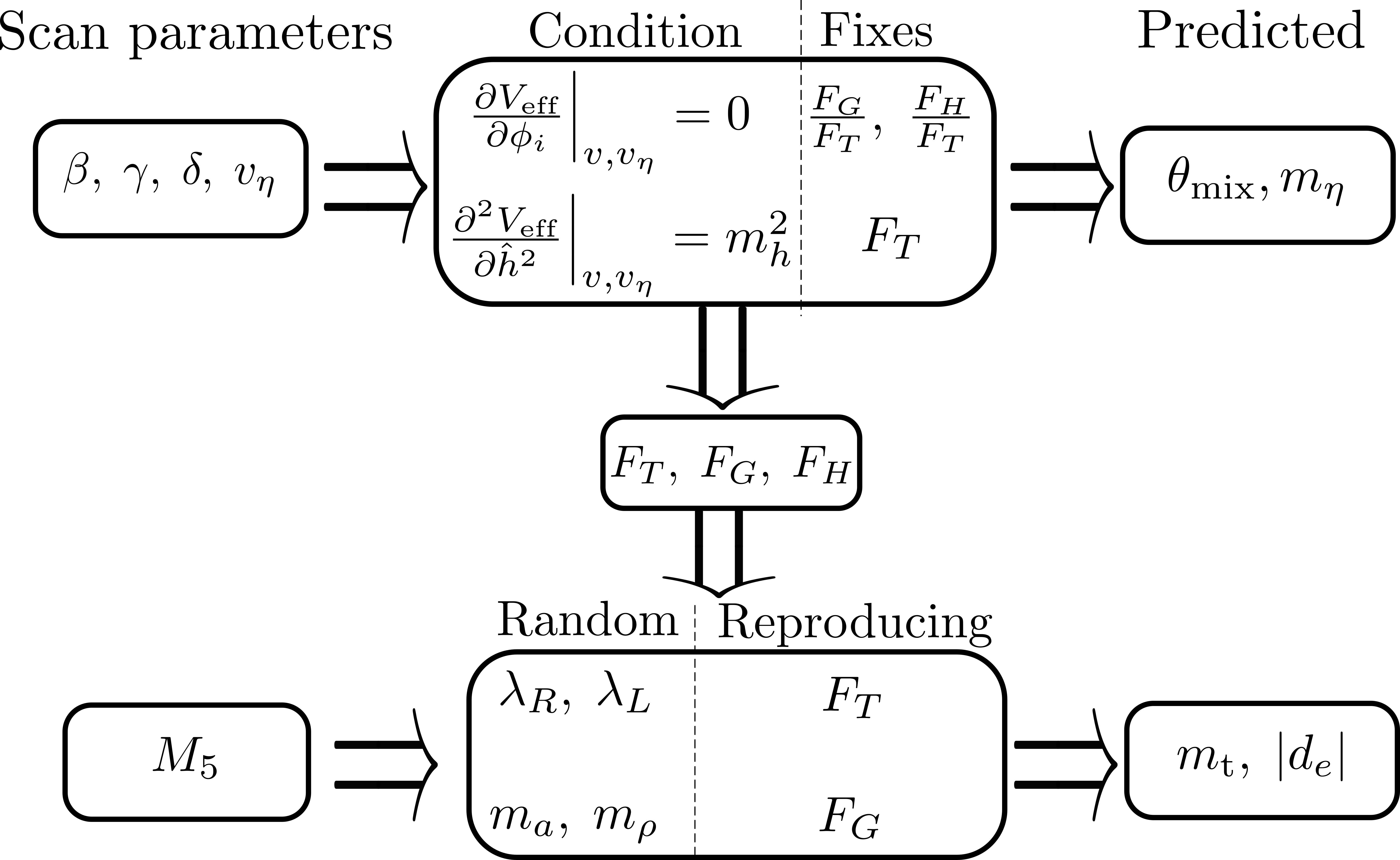}
    \caption{\sf\it Flowchart representation of the benchmarking procedure, see text for detailed explanations.}
    \label{fig:comp_sch}
\end{figure}

Before proceeding, we follow through the benchmark mining procedure, consistent with the current experimental constraints detailed above, see \cref{fig:comp_sch} for a flowchart representation. 
There are seven free parameters mentioned at the end of \cref{pot_zero}, namely $f$, $\beta,\gamma,\delta$, and $F_T,F_G,F_H$ which are involved in the pNGB potential. Among them we set $f = 1$ TeV, guided by the EW precision and Higgs couplings measurements at the LHC. There are two constraints arising from the requirement to reproduce the correct EW vev and the Higgs mass. We keep $v_\eta$ as a free parameter in the scan, to trade off a combination of other parameters. Thus, we initiate the parameter scan by selecting values for $v_{\eta}$ and the angles $\beta, \gamma, \delta$. Discrete symmetries of the potential allow us to restrict the scan to $\beta \in [0, \pi]$, $\gamma \in [0, \pi]$, $\delta \in [0, 2\pi]$, and $v_{\eta} > 0$.


To minimize the potential, we then solve for $F_G/F_T$ and $F_H/F_T$, while $F_T$ sets the overall scale of the potential, which are obtained by matching the EW vev, the Higgs mass, and ensuring that the scalar $\hat{h}$ is predominantly the doublet-like Higgs boson. Additionally, we compute emergent quantities like the singlet mass $m_{\eta} \geq 2 m_h $ (to prevent invisible Higgs decays), and the mixing angle $|\sin{\theta}| \lesssim 0.2$ (consistent with Higgs signal strength measurements).

Next, we determine the strong sector parameters, including the gauge resonance masses $m_{a,\rho}$ and fermionic masses $M_{L,R}$, compatible with the values of $F_G$ and $F_T$ obtained above. We further ensure that they satisfy $M_5 \geq 1$ TeV and $m_{\rho,a} \geq 1.5$ TeV, and reproduce the correct top quark mass. Finally, the electron EDM measurements limit the CP-violating couplings between the pNGBs and the top quark, restricting the angles $\gamma$ and $\beta$. In the next subsection, further constraints on the model parameters from the requirement of a strong FOPT are discussed.

\subsection{Gravitational waves from first-order phase transitions}
\label{strong FOPT}

Appearance of FOPTs across the model parameter space, which satisfy the phenomenological constraints discussed previously, are assessed by analyzing the thermal evolution of the finite temperature potential using the {\tt CosmoTransitions} code~\cite{Wainwright:2011kj}. A FOPT is feasible when the thermal evolution leads to two degenerate local minima separated by a potential barrier\footnote{We restrict the local minima within the range $|h|,|\eta| \leq f$, since the calculation of the Coleman-Weinberg potential does not take into account the effects of strong dynamics that may become relevant for large field values.}. 

\begin{figure}[t]
    \centering
    \includegraphics[width=0.5 \textwidth, keepaspectratio]{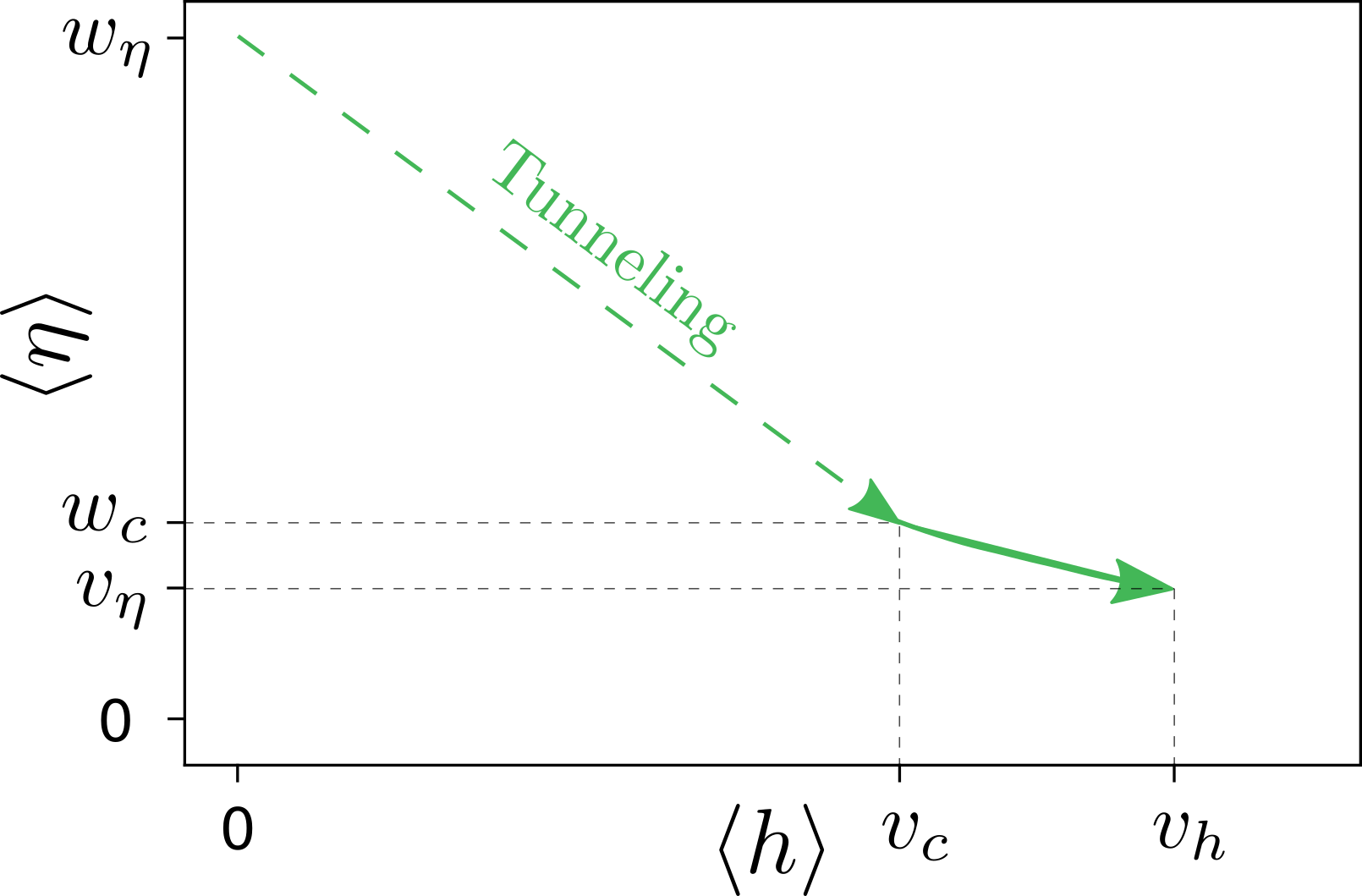}
    \caption{\sf\it Schematic plot showing the evolution of $h$ and $\eta$ expectation values as the  universe cools down.}
    \label{fig:transpath}
\end{figure}

In this model, the phase transition happens through a one-step pattern. We initiate the thermal evolution of the potential from a temperature $T_{\rm init} > f$, where the global minimum of the potential resides at $\left.(h,\eta)\right|_{T_{\rm init}} =(0,\omega_{\eta})$. As temperature decreases, a new local minimum emerges. Eventually the two existing minima become degenerate at a critical temperature ($T_c$) such that $V_{1\text{-loop}}(0,\omega_{\eta},T_c)=V_{1\text{-loop}}(v_c,\omega_{c},T_c)$. From this moment onwards, the system exhibits thermal tunneling, as shown in \cref{fig:transpath}. True vacuum bubbles start to form when the nucleation probability per unit time and volume becomes competitive with the Hubble expansion rate, see \cref{Thermo} for details.

Thermodynamic quantities, such as the nucleation temperature ($T_n$), inverse time duration of the transition ($\beta/H$), and latent heat released in the transition ($\alpha$) are calculated by solving the bounce equation \cite{Coleman:1977py,Callan:1977pt,Linde:1980tt}  (see \cref{Thermo} for definitions of the relevant thermodynamic quantities). Impact of higher dimensional operators involving derivative couplings of the pNGBs are neglected while solving the bounce equation, since they are suppressed by the scale $f\sim$ TeV. 

\begin{figure}[t!]
    \centering
    \includegraphics[width=0.6 \textwidth, keepaspectratio]{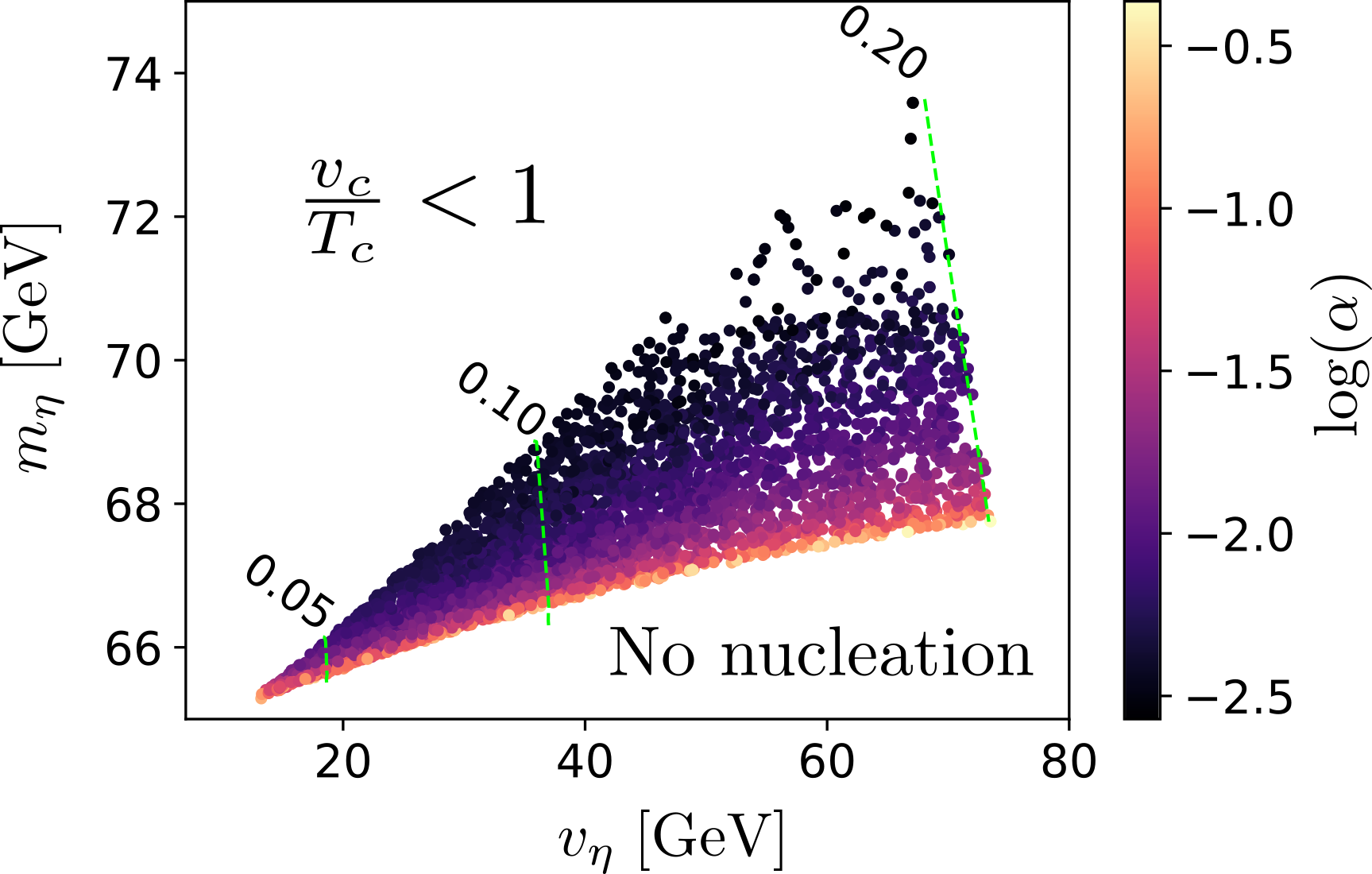}
    \caption{\sf\it Scatter plot of allowed parameter space with FOPTs in the $m_\eta-v_\eta$ plane for benchmarks with $\gamma=0$. The color coding depicts the logarithm of the latent heat $\alpha$ while the green-dashed contours show the mixing angle $\left| \sin \theta \right|$.}
    \label{fig:ms-vs}
\end{figure}

\begin{figure}[htbp]
    \centering
    \includegraphics[width=0.6 \textwidth, keepaspectratio]{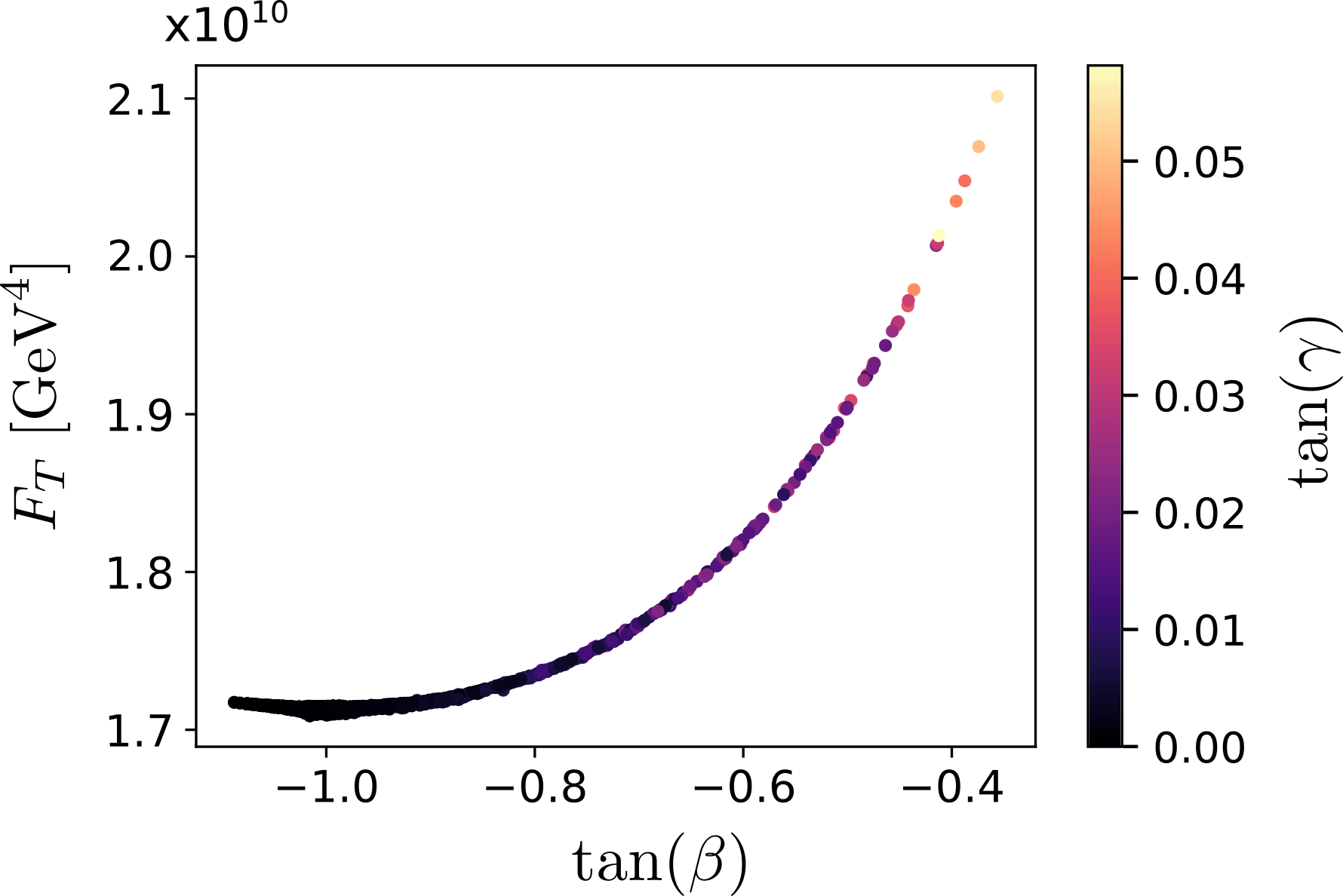}
    \caption{\sf\it Scatter plot showing the variation of $F_T$ with $\tan(\beta)$ for different values of $\tan(\gamma)$ which are allowed by the electron EDM constraints and which exhibit FOPT.}
    \label{fig:tanbeta-FT}
\end{figure}

\begin{figure}[t!]
    \centering
    \includegraphics[width=0.42 \textwidth, keepaspectratio]{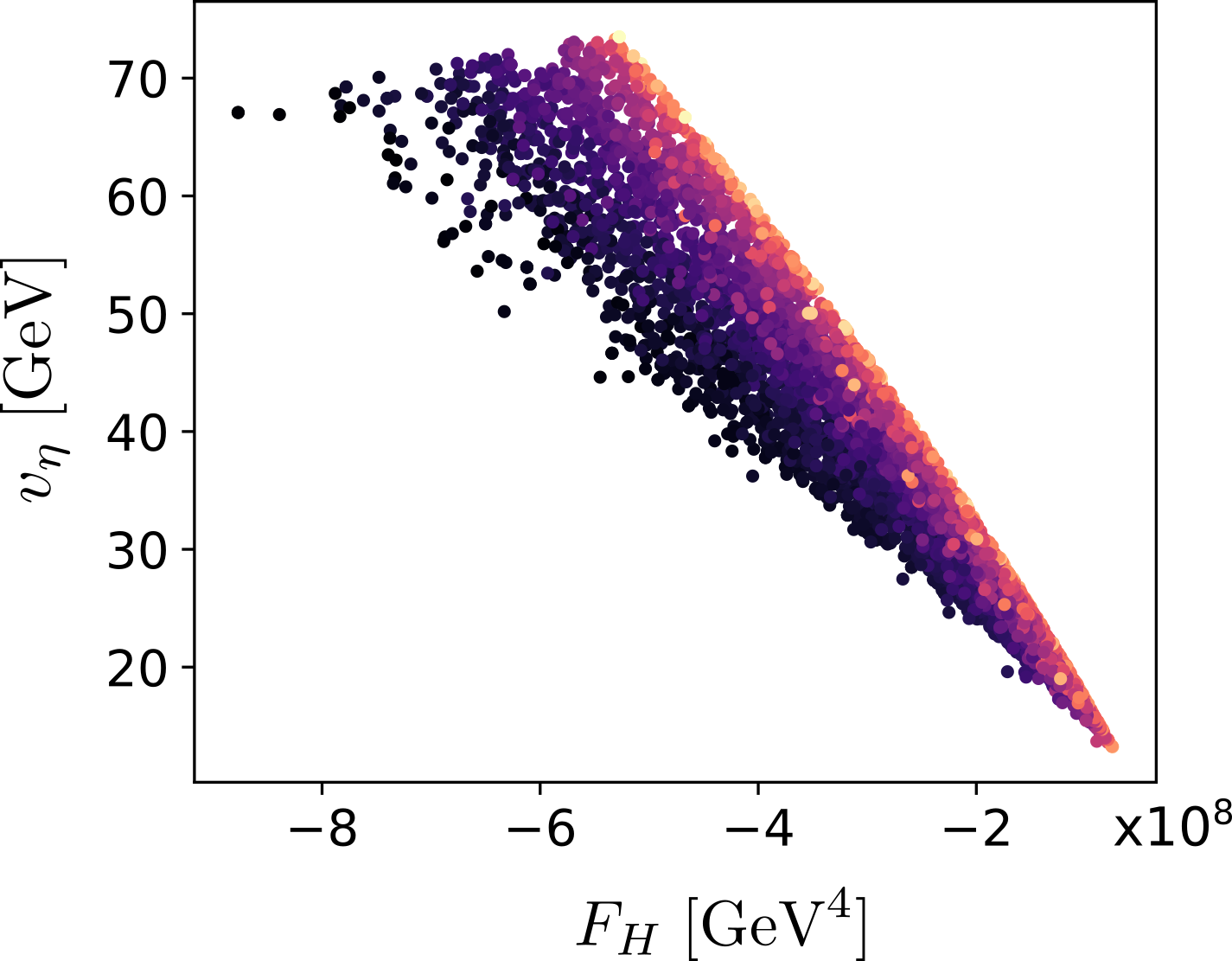}
    \hspace{5mm}
    \includegraphics[width=0.515 \textwidth, keepaspectratio]{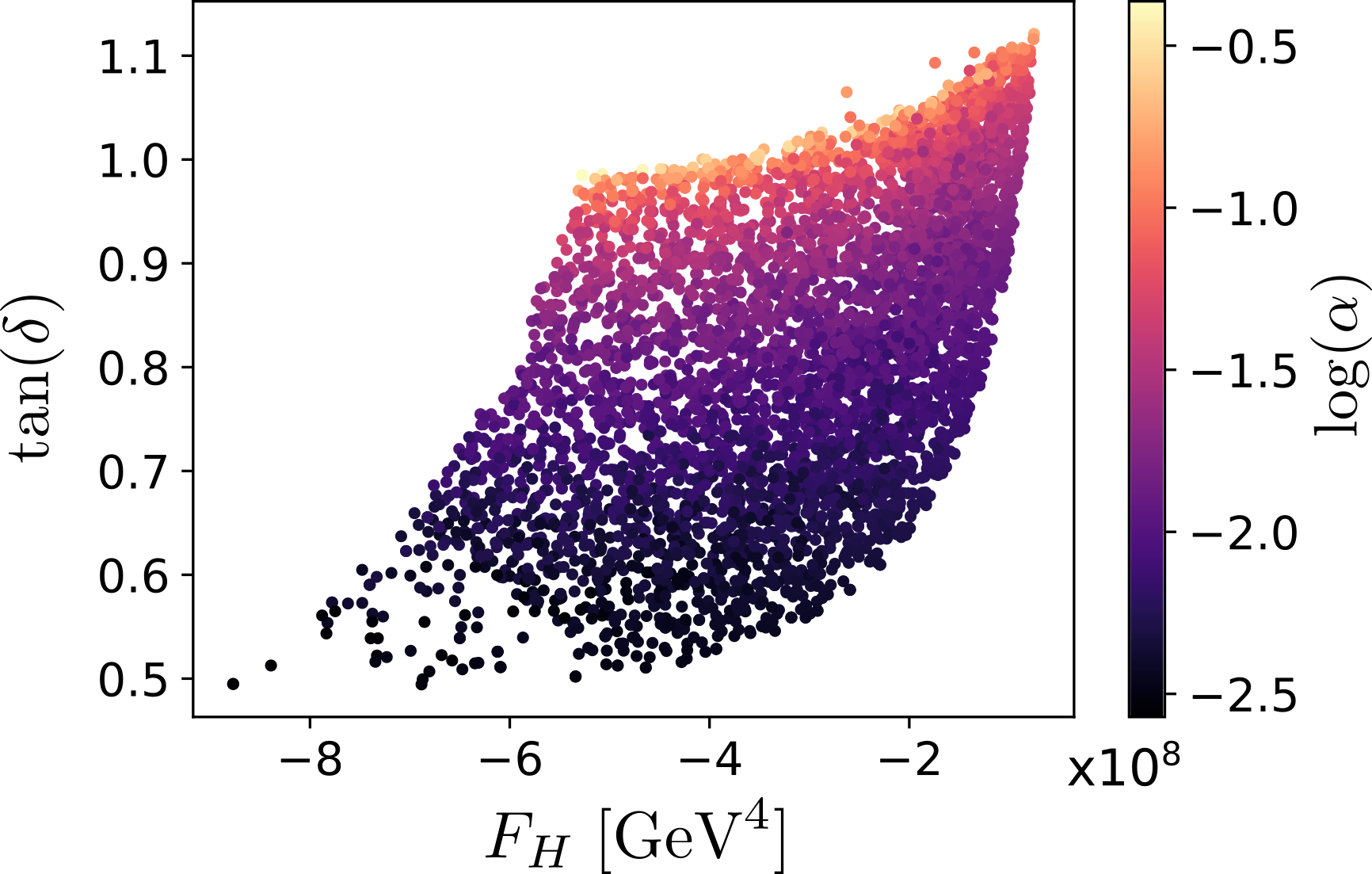}
    \caption{\sf\it Scatter plots of the nucleating benchmarks in the $v_\eta-F_H$ (left) and $\tan(\delta)-F_H$ (right) planes, varying $\tan(\gamma)\in[0,1]$. The color coding is same as \cref{fig:ms-vs}. 
   }
    \label{fig:FH_corr}
\end{figure}

The latent heat $\alpha$, characterising the strength of FOPT, vanishes for a smooth transition. Even though in such cases perturbative calculations may lead to a small non-zero value of $\alpha$, these results cannot be trusted as they usually lead to $v_c/T_c\ll 1$, where the perturbative computations break down. In contrast, we confine ourselves to the perturbative regime where $v_c/T_c>1$, yielding a strong FOPT (large values of $\alpha$), which in turn leads to strong GW signal. Additionally, we only consider the benchmark points where the bubbles of true vacuum nucleate. The requirement for nucleation restricts the parameter space to the case where $F_G \gg F_T \gg |F_H|$. We summarize the main results in \cref{fig:ms-vs,fig:tanbeta-FT,fig:FH_corr}. 

\cref{fig:ms-vs} shows that the requirement of a strong FOPT together with the constraints from \cref{LHC} prefers a lighter $\eta$ compared to the Higgs boson. The doublet-singlet mixing $|\sin\theta|$ (green-dashed lines) increases with $v_{\eta}$ and hence the upper bound $\left| \sin\theta \right|<0.2$ fixes the maximal $v_\eta$ value in the parameter scan. The region above the colored points, yields $v_c/T_c<1$, which is discarded from the perturbativity requirement. Whereas the region below the colored points, produces bounce solutions which do not satisfy the nucleation requirement. In \cref{fig:ms-vs} we fix $\gamma =0$, and observe that the strength of PT gradually drops with increasing mass of the pseudoscalar singlet for a fixed $v_\eta$. If one restricts to the points for which detectable GW signal is expected ($\log(\alpha)>-0.8$), a direct correlation connecting $m_\eta$ and $v_\eta$ is observed. For $\gamma \ne 0$, this correlation breaks, but large $m_\eta$ values are still disfavored by the strong transition requirement. 

We display in \cref{fig:tanbeta-FT}, a scatter plot between the variables $F_T$ and $\tan(\beta)$ with the colorbar showing the magnitude of the CP phase, $\tan(\gamma)$. We observe that all three variables are correlated, and large values of $\tan(\gamma)$ are not preferred when considering the current experimental constraints and the requirement of bubble nucleation. 

\cref{fig:FH_corr} shows how the strength of the FOPT varies in the $F_H-v_\eta$ and $F_H-\tan(\delta)$ planes. We present the benchmark points satisfying constraints from \cref{LHC}, varying $\tan(\gamma)\in [0,1]$. Clearly, larger values of $\tan(\delta)$ lead to stronger FOPTs, implying that the amount of CP violation is the main predicting factor for the strength of the FOPT. These figures underscore the significance of the hyperquark mass contribution to the scalar potential in determining the dynamics of the FOPT, despite being subleading compared to the contributions from the top quark and gauge sector. Notably, the vev of $\eta$ is primarily controlled by the tadpole term in $V_H$ (via $\mathcal{F}_H$, see \cref{LL_potential}), which depends on the phase $\delta$ and in turn dictates the dynamics of the PT. The phase $\gamma$ in $V_t$ (via $\mathcal{F}_t$), although contributes to $m_\eta$, has a comparatively minor role in the determining the strength of the PT. Therefore, unlike in \cref{fig:ms-vs}, large values of $\gamma$ do not significantly alter the correlations between $F_H-v_\eta$ and $F_H-\tan(\delta)$ that yield sufficiently strong transitions.


We emphasize that the parameters in the top quark sector, such as $M_5$, $\lambda^{L,R}$ enter into the potential (both at $T=0$ and finite temperature) via the combination $F_T$ (see \cref{LL_potential,F_coefficients}) and in the top mass through $F_t$ (\cref{mwmt}). Thus one can fix these parameters from the requirements of reproducing the correct value of the Higgs mass, small doublet-singlet mixing angle and the top quark mass at $T=0$. In fact, the $m_h=125$ GeV constraint impacts significantly the size of $F_T$. In contrast, the parameters in the gauge sector ($m_a,m_\rho$)  enjoy some freedom even after fixing the EW vev. We use the requirements of a strong FOPT to further constrain ($m_a,m_\rho$).

The strong FOPT is observed only within a narrow range of the parameter space, which may necessitate additional tuning beyond the typical fine-tuning present in composite Higgs models. However, a detailed quantitative assessment of fine-tuning in this model is beyond the scope of this paper. It is crucial to highlight that the strong FOPT offers complementary constraints compared to the existing experimental limits from the LHC and EDM measurements, substantially narrowing the permissible parameter range.

GWs can be sourced by the compression waves in the plasma and the bubble collisions, resulting from the FOPT. The latter source is relevant only if $\alpha \gg 1$. Since the transition strengths we found are moderately strong, i.e., $\alpha \sim \mathcal{O}(0.1)-\mathcal{O}(1)$, we only consider the GWs sourced by the compression waves in the plasma\footnote{GWs can also be sourced by turbulent motion in the plasma, however, this mechanism is not well-understood at present, see \cite{LISACosmologyWorkingGroup:2022jok, Roshan:2024qnv} and references therein for recent reviews.}. The GW spectrum sourced by sound waves in the plasma has been mapped out directly from numerical simulations \cite{Hindmarsh:2017gnf,Hindmarsh:2013xza,Hindmarsh:2015qta,Hindmarsh:2016lnk,Hindmarsh:2019phv} and the results have been digested into a phenomenological template which depends exclusively on the thermodynamic parameters of the transition as follows \cite{Caprini:2015zlo, Caprini:2019egz,LISACosmologyWorkingGroup:2022jok}
\begin{equation}
    \Omega_{\rm GW}h^2 = 4.13\times 10^{-7} \, \left(R_* H_*\right)  \left(1- \frac{1}{\sqrt{1+2\tau_{\rm sw}H_*}} \right)  \left(\frac{\kappa_{\rm sw} \,\alpha }{1+\alpha }\right)^2 \left(\frac{100}{g_*}\right)^\frac13 S_{\rm sw}(f).
\end{equation}
The spectral function of the sound wave and the peak frequency are given by 
\begin{equation}
 S_{\rm sw}(f)=\left(\frac{f}{f_{\rm sw}}\right)^3 \left[\frac{4}{7}+\frac{3}{7} \left(\frac{f}{f_{\rm sw}}\right)^2\right]^{-\frac72} \,, \quad f_{\rm sw} \,=2.6\times 10^{-5} {\rm Hz} \left(R_* H_*\right)^{-1} \left(\frac{T_*}{100 {\rm GeV}}\right)\left(\frac{g_*}{100}\right)^\frac16 \,.  
\end{equation}

The lifetime of the sound wave source ($\tau_{\rm sw}H_*$), marking the transition to a turbulence phase after the sound waves decay, is proportional to the bubble size and inversely proportional to the kinetic energy of the fluid. This can be expressed as
\begin{equation}
\tau_{\rm sw}H_* =\frac{R_*H_*}{U_f}\,, \quad U_f\approx \sqrt{\frac34 \frac{\alpha}{1+\alpha} \kappa_{\rm sw}}\,, \quad
R_*H_* \approx (8\pi)^\frac13 \, {\rm Max}(v_w,c_s)\left(\frac{\beta}{H}\right)^{-1} \, ,
\end{equation}
where $R_*H_*$ denotes the average bubble radius normalized to Hubble parameter, $U_f$ is the average velocity of the fluid, and $c_s$ is the speed of sound. The definitions of the inverse time duration of the transition $\beta/H$ and the wall velocity $v_w$ are given in \cref{Thermo}.
The quantity $\kappa_{\rm sw}$, parameterizing the amount of energy released that is transferred to the bulk kinetic motion of the fluid, is evaluated using the analytic fits given in \cite{Espinosa:2010hh} (see  also \cite{Giese:2020rtr} for recent developments). All the thermodynamic parameters used in the above expressions are evaluated at the nucleation temperature.

\begin{figure}[t]
    \centering
\includegraphics[width=0.48 \textwidth, keepaspectratio]{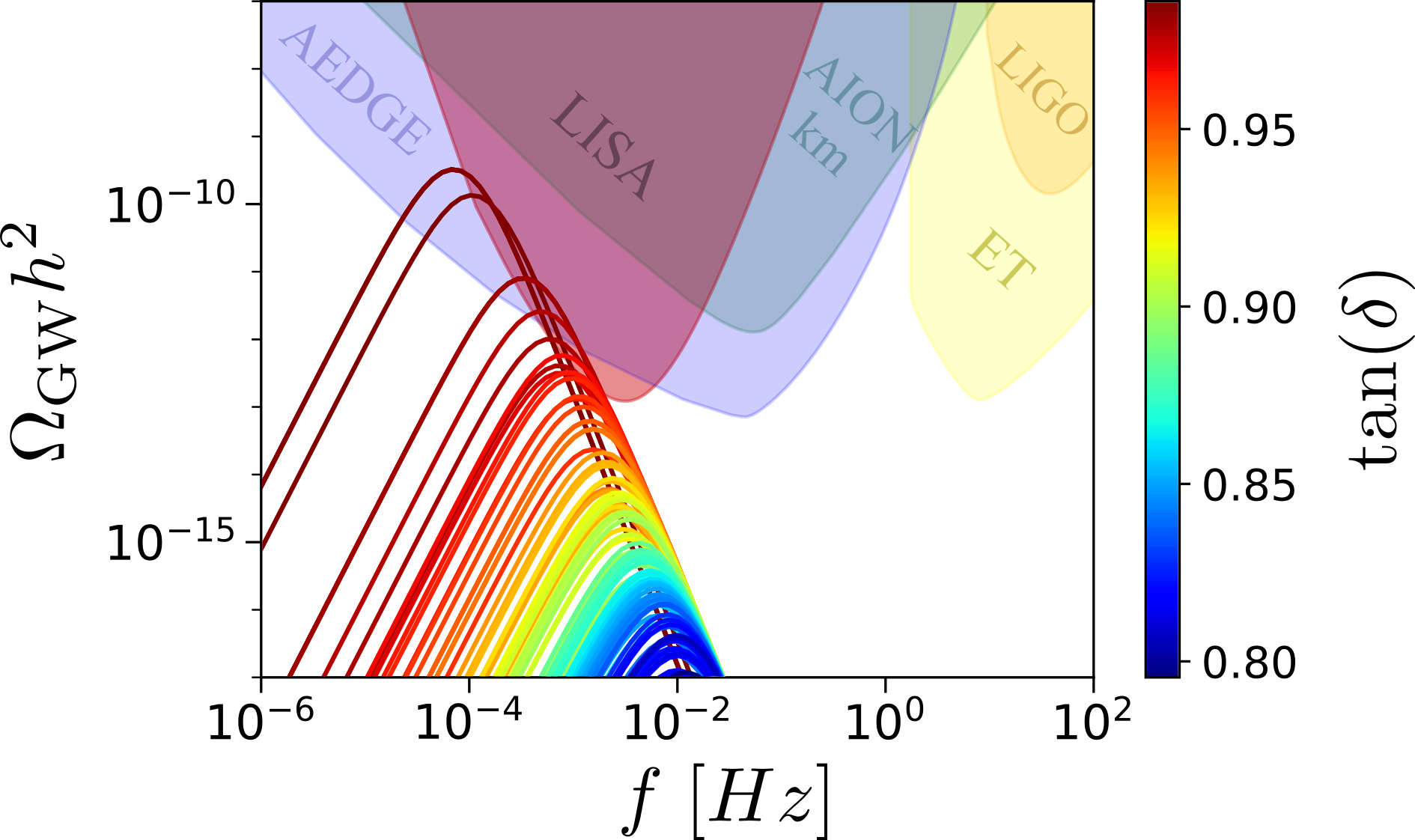}
        \hspace{5mm}
\includegraphics[width=0.475 \textwidth, keepaspectratio]{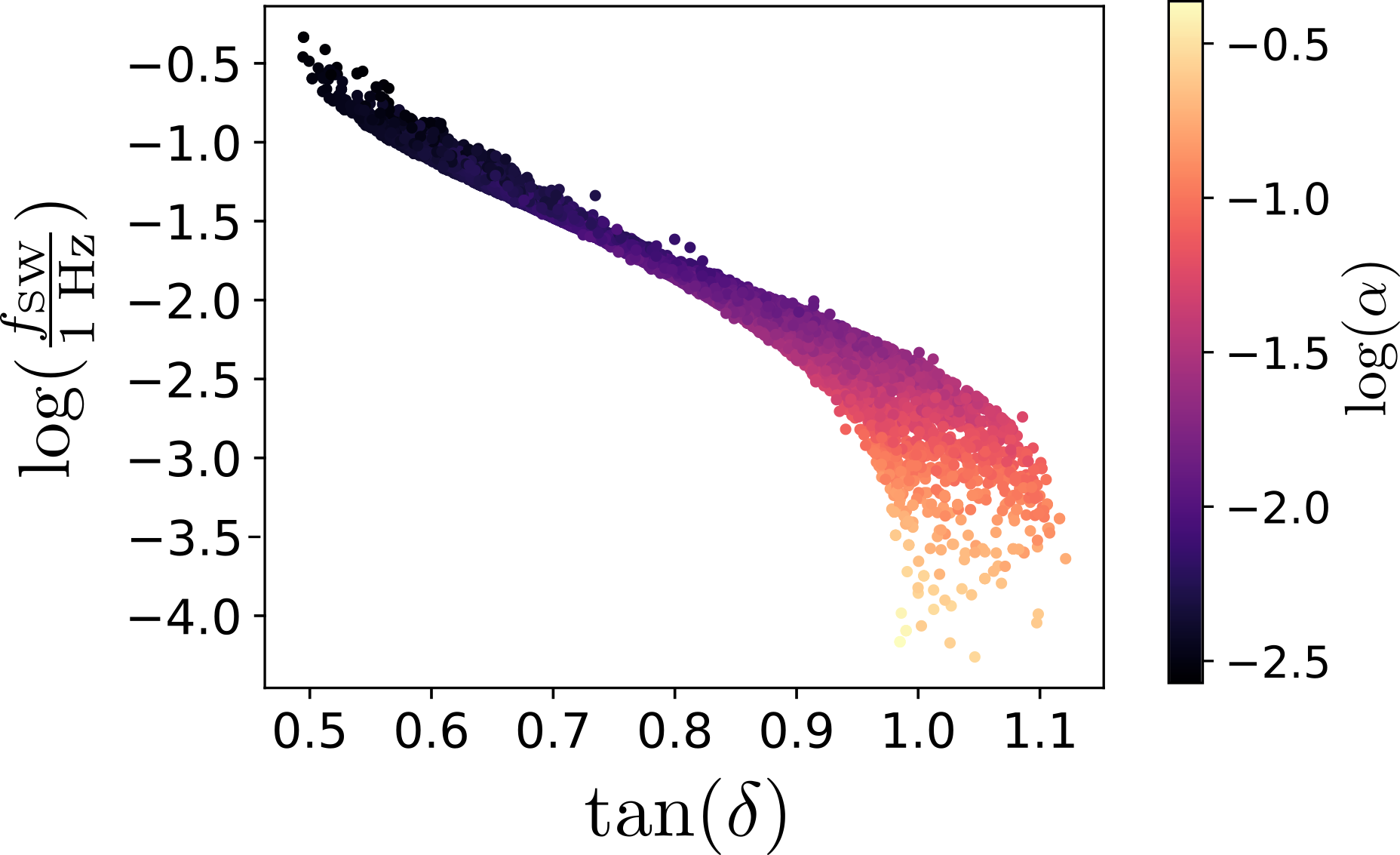}
    \caption{\sf\it 
    Left panel: predicted GW spectra obtained for benchmarks with $F_H=(5.25\pm0.25)\times10^{-3} {\rm ~TeV}^4$, combined with integrated sensitivity of the forthcoming GW detectors: AEDGE, LISA, AION 1km ET and LIGO (operational). 
    Right panel: Expected peak frequencies plotted against $\tan(\delta)$ for all the benchmarks. In both panels $\tan(\gamma)$ is varied in the range $[0,1]$.}
    \label{fig:GW_spectra}
\end{figure}

On the left panel of \cref{fig:GW_spectra}, we present predicted GW spectra for a subset of nucleating benchmarks within a small range of $F_H$, while varying $\delta$ and $\gamma$. Additionally, the integrated sensitivities of upcoming detectors, such as LISA \cite{2017arXiv170200786A}, AEDGE \cite{AEDGE:2019nxb}, AION 1km \cite{Badurina:2019hst}, ET \cite{Punturo:2010zz}, and the presently operational LIGO \cite{LIGOScientific:2016aoc,LIGOScientific:2016sjg,LIGOScientific:2017bnn,LIGOScientific:2017vox,LIGOScientific:2017ycc, LIGOScientific:2018mvr,LIGOScientific:2020ibl,LIGOScientific:2021usb}, are displayed. The maximal amplitude of the signal increases with CP violating phase $\tan(\delta)$ across the whole subset, up to the point beyond which the nucleation becomes impossible and thus no benchmarks are found. 

The right panel of \cref{fig:GW_spectra} illustrates the peak frequency dependence on $\tan(\delta)$, the key factor determining the strength of the PT, as discussed earlier. Amplitudes of the GW spectra that fall within the sensitivity range of AEDGE and LISA, resulting from strong FOPT ($\alpha\gtrsim 0.1$), require large value of $\tan(\delta) \sim 1$ and yield peak frequency of GWs in the range $10^{-3}-10^{-4}$ Hz.

\section{Conclusions}
\label{concl}

Composite Higgs models with underlying strongly coupled gauge theories with fermionic matter in the UV, that provide a resolution to the electroweak hierarchy problem, stand out as leading candidates to induce a FOPT in the early universe. The absence of any signatures for BSM physics at the LHC pushes the compositeness scale of a pNGB Higgs boson to the multi-TeV order, necessitating exploration of complementary avenues to probe the substructure of the Higgs boson. Upcoming gravitational wave detectors like LISA, AEDGE, among others, could provide new testing grounds for the cosmological phase transitions which in turn may shed some light on the origin of the Higgs potential and EW symmetry breaking. 

In this work we have investigated a maximally symmetric composite Higgs model with the next-to-minimal coset $SU(4)/Sp(4)$. The  presence of an extra pseudoscalar singlet in addition to the Higgs doublet in the field space manifold of this coset suggests a likely possibility for FOPT, prompting a detailed study of this coset. We have computed, in exact analytic form, the radiatively induced one-loop potential for the pNGBs at zero temperature, incorporating the effects of momentum-dependent form factors, and demonstrating the finiteness of the potential owing to the maximal symmetry.  We have verified that the exact expression for the potential, \cref{eq:V_ex}, matches the LL expansion in \cref{LL_potential} to a good approximation, and is suitable to use for practical purpose. The pNGB potential further receives a small contribution due to the hyeperquark masses, however, it plays a central role to determine the zero temperature vacuum. 

We have allowed for an explicit CP violation in the strong sector as well as in the partial compositeness mixing of the top quark, via introducing CP violating complex phases. This results in a pNGB potential at zero temperature which has a tadpole along the direction of the pseudoscalar singlet arising from the hyperquark mass contribution, and leads to a non-zero singlet vev at the minima of the potential. 

Using the imaginary time formalism, we have computed the finite temperature corrections to the one-loop potential in closed form and show that even in the presence of form factors, the thermal contributions become the standard expressions, that are usually found in theories with elementary scalars. The underlying reason lies in the fact that the contributions from the strong sector resonances to the thermal potential are exponentially suppressed if their masses significantly exceed the relevant temperature, which is around the electroweak scale in our case. 

Current bounds from the direct searches of new particles, measurements of EW observables and Higgs couplings at the LHC, as well as the limits on the CP violation from the measurement of the electron EDM are imposed to select the allowed parameter space of the model. Constraints from the electron EDM, presented in \cref{fig:tan(beta)-vs}, provide a correlation between $\beta$ and $\tan(\gamma)$, while the limit on the CP violation arising from $\delta$, which contributes to the EDM via the vev $v_\eta$ is relatively weaker.
By tracing the thermal evolution of the scalar potential as the universe cools down, we have identified the viable region among the allowed parameter space where a one-step FOPT occurs from a false vacuum to the electroweak symmetry breaking vacuum with a non-zero singlet vev, as illustrated in \cref{fig:transpath}. 

The requirement of strong FOPT further narrows the parameter space, establishing correlations between the strength of the FOPTs ($\alpha$) with the model parameters, such as the CP violating phase ($\gamma, \delta$), mass ($m_\eta$) and the vev ($v_\eta$) of the singlet etc, as illustrated in \cref{fig:ms-vs,fig:tanbeta-FT,fig:FH_corr}. Specifically, we observe that a large CP violation arising from the hyperquark masses through $\tan(\delta)$ is required for strong transitions, which is weakly constrained from the EDM measurements. The gravitational waves emitted from the sound waves sourced by strong FOPTs lie within the detectable frequency and sensitivity range of upcoming experiments, particularly LISA and AEDGE (see \cref{fig:GW_spectra}), thus offering alternate avenues to probe the maximally symmetric composite Higgs model.

\section*{Acknowledgments}

We would like to thank Anish Ghoshal for collaboration during the early stages of this work. M.M. and I.N. thank Bogumi\l{}a \'Swie\.zewska for helpful discussions. A.B. is supported by the Department of Atomic Energy, Govt.\ of India. A.B. also acknowledges support from the Knut and Alice Wallenberg foundation under the grant KAW 2017.0100 (SHIFT project), and the Chalmers University of Technology, G\"oteborg, Sweden during the initial stages of this work. A.B. would like to express special thanks to the Mainz Institute for Theoretical Physics (MITP) of the Cluster of Excellence PRISMA$^+$ (Project ID 390831469), for its hospitality and partial support during the completion of this work. M.M. received support from Norwegian Financial Mechanism for years 2014-2021, grant nr DEC-2019/34/H/ST2/00707. M.M. acknowledges support by the Swedish Research Council (Vetenskapsr{\aa}det) through Contract No. 2017-03934. IN was partially supported by the National Science Centre, Poland, under research grant no. 2020/38/E/ST2/00243.

\appendix

\section{Model details}
\label{model_details}

The building blocks for constructing the low energy Lagrangian of the $SU(4)/Sp(4)$ coset are discussed in \cite{Banerjee:2022izw}. For completion, here we present some of the relevant details. 

\subsection*{Generators}
\label{generators}
The $SU(4)$ generators are divided into three categories, six unbroken generators spanning the $SU(2)_L\times SU(2)_R$ subgroup of $Sp(4)$, remaining four unbroken $Sp(4)$ generators and five broken generators of the $SU(4)/Sp(4)$ coset. The generators spanning the $SU(2)_L\times SU(2)_R$ subgroup are
\begin{small}
\begin{align}
T^{1}_L=\frac{1}{2}\left(
\begin{array}{cccc}
 \sigma^1 & 0 \\
 0 & 0 
\end{array}
\right),
T^{2}_L=\frac{1}{2}\left(
\begin{array}{cccc}
 \sigma^2 & 0 \\
 0 & 0 
\end{array}
\right),
T^{3}_L=\frac{1}{2}\left(
\begin{array}{cccc}
 \sigma^3 & 0 \\
 0 & 0 
\end{array}
\right), \\
T^{1}_R=\frac{1}{2}\left(
\begin{array}{cccc}
 0 & 0 \\
 0 & \sigma^1 
\end{array}
\right),
T^{2}_R=\frac{1}{2}\left(
\begin{array}{cccc}
 0 & 0 \\
 0 & \sigma^2 
\end{array}
\right),
T^{3}_R=\frac{1}{2}\left(
\begin{array}{cccc}
 0 & 0 \\
 0 & \sigma^3 
\end{array}
\right).
\end{align}
\end{small}
The rest of the $Sp(4)$ generators are
\begin{small}
\begin{align}
   T^{7}=\frac{1}{2\sqrt{2}}\left(
\begin{array}{cccc}
 0 & \mathbb{I} \\
 \mathbb{I} & 0
\end{array}
\right),
 T^{8}=\frac{i}{2\sqrt{2}}\left(
\begin{array}{cccc}
 0 & -\sigma^1 \\
 \sigma^1 & 0 
\end{array}
\right),
 T^{9}=\frac{i}{2\sqrt{2}}\left(
\begin{array}{cccc}
 0 & -\sigma^2 \\
 \sigma^2 & 0 
\end{array}
\right),
 T^{10}=\frac{i}{2\sqrt{2}}\left(
\begin{array}{cccc}
 0 & -\sigma^3 \\
 \sigma^3 & 0 
\end{array}
\right),
\end{align}
\end{small}
and the broken generators are given by
\begin{small}
\begin{align}
   T^{\hat{1}}=\frac{1}{2}\left(
\begin{array}{cccc}
 0 & \sigma^1 \\
 \sigma^1 & 0 
\end{array}
\right),
T^{\hat{2}}=\frac{1}{2}\left(
\begin{array}{cccc}
 0 & \sigma^2 \\
 \sigma^2 & 0 
\end{array}
\right),
T^{\hat{3}}=\frac{1}{2}\left(
\begin{array}{cccc}
 0 & \sigma^3 \\
 \sigma^3 & 0 
\end{array}
\right),
T^{\hat{4}}=\frac{i}{2}\left(
\begin{array}{cccc}
 0 & -\mathbb{I} \\
 \mathbb{I} & 0 
\end{array}
\right),
T^{\hat{5}}=\frac{1}{2}\left(
\begin{array}{cccc}
 \mathbb{I} & 0 \\
 0 & -\mathbb{I} 
\end{array}
\right).
\end{align}
\end{small}

\subsection*{Kinetic mixing between the pNGBs}
\label{kinetic_mixing}

In the unitary gauge, expression for the non-linearly transforming pNGB matrix ($\Sigma$) is given by 
\begin{align}\label{eq:def_Sigma}
   \Sigma = \left(\begin{array}{cccc}
        c_{\pi_0/2} + i \frac{\eta}{\pi_0}s_{\pi_0/2} & 0 & -\frac{h}{\pi_0}s_{\pi_0/2} & 0\\
        0 &c_{\pi_0/2} + i \frac{\eta}{\pi_0}s_{\pi_0/2} & 0 & -\frac{h}{\pi_0}s_{\pi_0/2}\\
        \frac{h}{\pi_0}s_{\pi_0/2} & 0 & c_{\pi_0/2} - i \frac{\eta}{\pi_0}s_{\pi_0/2} & 0\\
        0 & \frac{h}{\pi_0}s_{\pi_0/2} & 0 & c_{\pi_0/2} - i \frac{\eta}{\pi_0}s_{\pi_0/2}
    \end{array}\right). 
\end{align}
The kinetic term for the pNGBs are written using $\Sigma$ as
\begin{equation}
\mathcal{L}_{\text{kinetic}}   \equiv \frac{f^2}{8} \text{Tr} |D_{\mu} \Sigma|^2  \supseteq \frac{1}{2}g_{\hat{a}\hat{b}}(\pi) \partial_{\mu}\pi^{\hat{a}} \partial^{\mu} \pi^{\hat{b}}, \quad {\rm where} \quad \pi^{\hat{a}} = (h, \eta),
\end{equation}
and the field-space metric is given by 
\begin{equation}
    g_{\hat{a}\hat{b}}(\pi) \equiv  \frac{f^2}{\pi_0^2} s_{\pi_0}^2  \left[ \delta_{\hat{a}\hat{b}} + \frac{\pi_{\hat{a}} \pi_{\hat{b}}}{\pi_0^2} \left( \left( \frac{\pi_0}{f} \right)^2 \frac{1}{s_{\pi_0}^2 }-1 \right) \right]. \label{metric} 
\end{equation}
When both $h$ and $\eta$ receive vevs, the field space metric leads to kinetic mixing between the two scalar fields. The non-trivial geometry induced by the field space metric can be locally made flat around the vacuum, by a non-unitary transformation on the vacuum metric $g_{\hat{a}\hat{b}}(v,v_\eta)$. We expand the kinetic term around the vacuum and use the transformations $h = a h^\prime$, $\eta = b h^\prime + c \eta^\prime$, where $h^\prime$ and $\eta^\prime$ are the canonically normalized fields. The expressions for the coefficients $(a,b,c)$ are found to be 
\begin{small}
\begin{equation}
    a = \frac{1}{\langle s_{\pi_0}\rangle} \sqrt{ \frac{v_\eta^2}{f^2}  + \frac{v^2 \langle s_{\pi_0}\rangle^2  }{\langle\pi_0\rangle^2}  },~
    b=  \frac{ \sqrt{2} v v_\eta  }{f \langle\pi_0\rangle }  \frac{ \left(  f^2 
 - \langle\pi_0\rangle^2 /\langle s_{\pi_0}\rangle^2 \right)  \langle s_{\pi_0}\rangle }{ \sqrt{f^2 v^2 + 2v_\eta^2 \langle\pi_0\rangle^2  - f^2 v^2 \langle c_{2\pi_0}\rangle  }},~
 c = \frac{  \langle\pi_0\rangle^2 }{ \sqrt{ f^2 v^2 \langle s_{\pi_0}\rangle^2 + v_\eta^2 \langle\pi_0\rangle^2 } },
    \label{eq:c}
\end{equation}
\end{small}
where quantities appearing inside $\langle\cdots\rangle$ brackets indicate that the fields are evaluated at their vevs.

\subsection*{Spurions and form factors}
\label{form__factors}

The spurions appearing in \cref{top_lag} which determine how the elementary quarks are embedded in the anti-symmetric $\mathbf{6}$ of $SU(4)$ are given below:
\begin{small}
\begin{align}
    A^t_{L}= \frac{i}{2\sqrt{2}}\left(
\begin{array}{cccc}
 0 & \sigma^2-i\sigma^1 \\
 \sigma^2+i\sigma^1 & 0 
\end{array}
\right),~
    A^b_{L}= \frac{1}{2\sqrt{2}}\left(
\begin{array}{cccc}
 0 & \mathbb{I} -\sigma^3 \\
 -\mathbb{I}+\sigma^3 & 0 
\end{array}
\right),~  
    A^t_{R}= \frac{i}{\sqrt{2}}\left(
\begin{array}{cccc}
 c_\beta\sigma^2 & 0 \\
 0 & e^{i\gamma}s_\beta \sigma^2 
\end{array}
\right).
\end{align}    
\end{small}
Note that there are two options for embedding the right-handed top quark, stemming from the decomposition of $\mathbf{6}$ of $SU(4)$ under $SU(2)_L\times SU(2)_R$ as $\mathbf{6}\to (\mathbf{2},\mathbf{2})+2\cdot(\mathbf{1},\mathbf{1})$, which yields two $(\mathbf{1},\mathbf{1})$ representations suitable for accommodating $t_R$.

The expressions for the form factors given in \cref{top_lag}, which encapsulate the strong dynamics of the top-partners ($\Psi_{5,1}$) are shown in terms of Euclidean momentum as \cite{Dong:2020eqy}
\begin{align}
    &\Pi^{L,R}_0=\Pi^{L,R}_{\Psi_5} =\frac{|\lambda^{L,R}_{5}|^2 f^2}{p^2+M^2_{5}}, 
    & &\Pi^{L,R}_1=\Pi^{L,R}_{\Psi_1} -\Pi^{L,R}_{\Psi_5} =\frac{|\lambda^{L,R}_{1}|^2 f^2}{p^2+M^2_{1}} - \frac{|\lambda^{L,R}_{5}|^2 f^2}{p^2+M^2_{5}}, \\
    &\Pi^{LR}_2=\Pi^{LR}_{\Psi_5} = \frac{\lambda^{L}_{5}\lambda^{R*}_{5} f^2 M_5}{p^2+M^2_{5}},
    & &\Pi^{LR}_1=\Pi^{LR}_{\Psi_1} -\Pi^{LR}_{\Psi_5} = \frac{\lambda^{L}_{1}\lambda^{R*}_{1} f^2 M_1}{p^2+M^2_{1}} - \frac{\lambda^{L}_{5}\lambda^{R*}_{5} f^2 M_5}{p^2+M^2_{5}}.
\end{align}
In the maximally symmetric case $M_1=M_5$ and $\lambda^{L,R}_1=\lambda^{L,R}_5$, which implies $\Pi^{L,R}_1=0$ and $\Pi^{LR}_1=0$.

The form factors $\Pi^{W,B}_0$ in the gauge sector are given by \cite{Contino:2010rs}
\begin{align}
    \Pi^W_0 = \frac{p^2}{g^2}\left(1+ \frac{g^2 f_\rho^2}{p^2+m_\rho^2} \right) \simeq \frac{p^2}{g^2}\,, \quad \Pi^B_0 = \frac{p^2}{g^{\prime 2}}\left(1+ \frac{g^{\prime 2} f_\rho^2}{p^2+m_\rho^2} \right) \simeq \frac{p^2}{g^{\prime 2}}\,,  \label{approximation_1}
\end{align}    
while $\Pi_1(p^2)$ is given by
\begin{align}
    \Pi_1(p^2) = f^2 + 2p^2 \left(\frac{f_a^2}{p^2+m_a^2} - \frac{f_\rho^2}{p^2+m_\rho^2}\right).
    \label{eq:Pi_1_g_form}
\end{align}
Using the Weinberg's sum rules, shown in \cref{sumrules}, we obtain
\begin{align}
    \Pi_1(p^2) = \frac{f^2 m_a^2 m_\rho^2}{(p^2+m_a^2)(p^2+m_\rho^2)}. 
    \label{eq:Pi_1_g_WSR}
\end{align}

\section{Details on exact computation of thermal potential}
\label{app:Viett}

Some additional technical details regarding the exact analytic computation of the thermal potential, as shown in \cref{pot_thermal} are discussed below.



\subsection{Exact expression for zero temperature potential}

First, we show that the zero temperature part of the thermal potential separated in \eqref{eq:V^T_genIV}, and given by
\begin{equation}
    V^{(T=0)}_{\text{CW}}(\tilde{m}^2_i)=\frac{N_\text{eff}}{2}\int \frac{d^3p}{(2\pi)^3}\left[\sum^3_{i=1} \tilde{E}_i-|\vec{p}|-E_a-E_\rho\right],
    \label{eq:VCWex}
\end{equation}
is equivalent to \cref{eq:V^T_gen_1}, where we restore the field independent terms from \cref{eq:V^T_genII}. With the help of the following identity~\cite{Dolan:1973qd}\footnote{Here we use Euclidean momenta, which causes a factor of $i$ difference between~\cite{Dolan:1973qd} and our expressions for $V^{(T=0)}_{\text{CW}}$.}
\begin{equation*}
    \int \frac{dp_0}{2\pi}\log(p_0^2+|\vec{p}|^2+m^2-i\epsilon)=\sqrt{|\vec{p}|^2+m^2},
\end{equation*}
the potential in \cref{eq:VCWex} takes the form
\begin{equation}
V^{(T=0)}_{\text{CW}}(\tilde{m}^2_i) =    \frac{N_\text{eff}}{2}\int \frac{d^4p}{(2\pi)^4}\log\left[\frac{(\tilde{m}^2_1+p^2) (\tilde{m}^2_2+p^2) (\tilde{m}^2_3+p^2)}{p^2 E_a^2 E_\rho^2} \right].
\end{equation}
Finally using the Vi\`ete equations \eqref{eq:ViettIII}, we recover \cref{eq:V^T_gen_1} with $m_{1,2}\equiv m_{\rho,a}$ as
\begin{equation}
      V^{(T=0)}_{\text{CW}}(\tilde{m}^2_i) =  \frac{N_\text{eff}}{2}\int \frac{d^4p}{(2\pi)^4}\log\left(1+\frac{m^2_\text{SM} m^2_a m^2_\rho}{p^2(p^2+m_a^2)(p^2+m_\rho^2)}\right).
\end{equation}
Thus, we prove that \eqref{eq:V_ex} is indeed an equivalent expression for the one-loop zero temperature potential. Now, we evaluate the integral \eqref{eq:VCWex} to obtain an exact analytic expression for the zero temperature Coleman-Weinberg potential. Since, each term in the potential \eqref{eq:VCWex} has the same momentum dependence, the only integral we have to compute is
\begin{equation}
    \int\frac{d^3p}{(2\pi)^3}\frac{E_i}{2}= \frac{1}{2\pi^2}\lim_{\Lambda\to\infty}\int^\Lambda_0 dp\, |\vec{p}|^2 \frac{\sqrt{|\vec{p}|^2+m_i^2}}{2},
    \label{eq:limit}
\end{equation}
with $E_i\equiv \sqrt{|\vec{p}|^2+m_i^2}$. In the last equality we include a cut-off $\Lambda$ to regulate the UV behaviour of the integral. In what follows, we will show that \cref{eq:VCWex} is independent of this cut-off and the final result is fully convergent. Performing the integral in \cref{eq:limit} we obtain
\begin{align}
    \int\frac{d^3p}{(2\pi)^3}\frac{E_i}{2} &=\frac{1}{32\pi^2}\lim_{\Lambda\to\infty}\left[\Lambda\sqrt{\Lambda^2+m_i^2}(m_i^2+2\Lambda^2)-m_i^4\log\left(\frac{\Lambda}{m_i}+\sqrt{1+\frac{\Lambda^2}{m_i^2}}\right)\right].
\end{align}
Expanding around $m_i/\Lambda\to 0$ and keeping only positive powers of $\Lambda$, \eqref{eq:VCWex} becomes
\begin{align}
        \nonumber 
        V^{(T=0)}_{\text{CW}}(\tilde{m}^2_i) &= \frac{N_{\rm eff}}{32\pi^2}\left[\sum_{i=1}^3\tilde{m}^4_i\log\left(\frac{\tilde{m}_i}{\mu}\right)-m^4_a\log\left(\frac{m_a}{\mu}\right)-m^4_\rho\log\left(\frac{m_\rho}{\mu}\right) + \frac{1}{4}\Big(\sum^3_{i=1}\tilde{m}^4_i-m_a^4-m_\rho^4\Big)\right] \\[2mm]
        &+\frac{N_{\rm eff}}{32\pi^2}\lim_{\Lambda\to\infty}\left[2\Lambda^2\Big(\sum^3_{i=1}\tilde{m}^2_i-m_a^2-m_\rho^2\Big)-\Big(\sum^3_{i=1}\tilde{m}^4_i-m_a^4-m_\rho^4\Big)\log\left(\frac{2\Lambda}{\mu}\right)\right].
        \label{eq:V_CWfull}
\end{align}
where we introduce an arbitrary scale $\mu$ to keep the arguments of the logarithms dimensionless, while separating the $\Lambda$ dependent divergent terms from the finite ones. Several observations are worth noting regarding this expression:
\begin{itemize}
    \item The $\Lambda^4$ terms completely drop from the \cref{eq:V_CWfull}.
    \item The $\Lambda^2$ terms cancel, after substituting $\tilde{m}^2_1 +\tilde{m}^2_2 +\tilde{m}^2_3= m_a^2+ m_\rho^2$, (see \cref{eq:ViettIII}).
    \item Using the Vi\`ete equations \eqref{eq:ViettIII}, one can write
    \begin{align}
    \tilde{m}^4_1+\tilde{m}^4_2+\tilde{m}^4_3=(\tilde{m}^2_1+\tilde{m}^2_2+\tilde{m}^2_3)^2-2(\tilde{m}^2_1 \tilde{m}^2_2+\tilde{m}^2_1 \tilde{m}^2_3+\tilde{m}^2_2\tilde{m}^2_3)=m_a^4+m_\rho^4.
    \label{no_log_div}
    \end{align} 
    Thus, the prefactor of the logarithmic divergent term also vanishes, rendering the potential fully finite, as expected in the maximally symmetric case.  
\end{itemize}
For clarity, we give the final convergent expression for the zero temperature potential as
\begin{align}
        V^{(T=0)}_{\text{CW}}(\tilde{m}^2_i) &= \frac{N_{\rm eff}}{32\pi^2}\left[\sum_{i=1}^3\tilde{m}^4_i\log\left(\frac{\tilde{m}_i}{\mu}\right)-m^4_a\log\left(\frac{m_a}{\mu}\right)-m^4_\rho\log\left(\frac{m_\rho}{\mu}\right)\right].
        \label{eq:VCWFULL}
\end{align}
Despite appearances, the field dependent terms in the above potential are insensitive to specific choice of the scale $\mu$. Under an arbitrary shift $\mu\to \mu+\Delta\mu$, the field-dependent terms in \cref{eq:VCWFULL} change as  
\begin{equation}
   \sum_{i=1}^3\tilde{m}^4_i\log\left(\frac{\tilde{m}_i}{\mu}\right) \to \sum_{i=1}^3\tilde{m}^4_i\left[\log\left(\frac{\tilde{m}_i}{\mu}\right)+\log\left(\frac{\mu}{\mu+\Delta\mu}\right)\right].
\end{equation}
Using the relation $\tilde{m}^4_1+\tilde{m}^4_2+\tilde{m}^4_3=m_a^4+m_\rho^4$ (see \cref{no_log_div}), we find that the shifted term in the above equation is field independent, proving that the choice of $\mu$ does not impact the field-dependent part of the scalar potential.
\subsection{Generalization to other composite models}\label{app:V_gen}

The procedure used to derive the exact form of the one-loop potential at zero and finite temperature is also applicable to other composite models, although the result will not be necessarily finite. At zero temperature, a generic contribution to the radiatively generated pNGB potential in a composite Higgs scenario is given by
\begin{equation}
V^{\text{(gen)}}_{\text{eff}}=\frac{N_{\text{eff}}}{2}\int \frac{d^4p}{(2\pi)^4} \log\left[1+\frac{\Pi_1(p^2,\pi^{\hat{a}})}{\Pi_2(p^2,\pi^{\hat{a}})} \right]\,,
\label{V_gen_CH_0}
\end{equation}
where $N_{\rm eff}$ denotes an overall field-independent prefactor, and the form factors $\Pi_{1,2}(p^2,\pi^{\hat{a}})$ are generally parameterized as
\begin{align}
    \Pi_{1,2}(p^2,\pi^{\hat{a}}) = \sum_{i} \frac{A^{(i)}_{1,2}(\pi^{\hat{a}})}{p^2+m_i^2}\,.
\end{align}
Here $A^{(i)}_{1,2}(\pi^{\hat{a}})$ are model dependent functions of the pNGB fields $\pi^{\hat{a}}$ with appropriate dimension, and $m_i$ are the masses of the composite resonances. Typically, minimal number of resonances with lightest masses, which render the momentum integral finite, are taken into account in the summation. However, the calculation we present below is independent of the finiteness of the potential. The generic potential in \cref{V_gen_CH_0} can be brought to the form
\begin{equation}
V^{\text{(gen)}}_{\text{eff}}=\frac{N_{\text{eff}}}{2}\int \frac{d^4p}{(2\pi)^4} \log\left[\frac{G_r(p^2,\pi^{\hat{a}})}{G_q(p^2,\pi^{\hat{a}})} \right],
\label{V_gen_CH}
\end{equation}
where $G_r$ and $G_q$ are some polynomials in $p^2$ of order $r$ and $q$, respectively, whose coefficients depend on resonance masses and pNGB fields $\pi^{\hat{a}}$. Employing the factorizability of a polynomial over complex space we rewrite \cref{V_gen_CH} with factorized numerator and denominator
\begin{equation}
V^{\text{(gen)}}_{\text{eff}}=\frac{N_{\text{eff}}}{2}\int \frac{d^4p}{(2\pi)^4} \log\left[\frac{ (p^2+\tilde{m}^2_1)(p^2+\tilde{m}^2_2)...(p^2+\tilde{m}^2_r)}{(p^2+\tilde{m}^2_{r+1})(p^2+\tilde{m}^2_{r+2})...(p^2+\tilde{m}^2_{r+q})} \right],
\label{V_gen_CH_fact}
\end{equation}
To find $\tilde{m}^2_i$, which are in general field-dependent, we  compare the polynomial coefficients of $p^2$ in $G_{r\;(q)}(p^2,\pi^{\hat{a}})$ to that of the numerator (denominator) in \cref{V_gen_CH_fact}. This yields $r$ equations for parameters $\tilde{m}_1,\;\tilde{m}_2,...,\tilde{m}_{r}$ that appear in the numerator and $q$ equations for parameters $\tilde{m}_{r+1},\;\tilde{m}_{r+2},...,\tilde{m}_{r+q}$ appearing in the denominator.

 Next, we use the imaginary-time formalism to get the expression for the potential at zero and finite temperature. After substituting \cref{Imaginary_formalism}, the integral \eqref{V_gen_CH_fact} becomes
\begin{equation}
 V^{\text{(gen)}}_{\text{eff}}=\frac{N_{\text{eff}}}{2}T\sum_{n\in\mathbb{Z}}\int \frac{d^3p}{(2\pi)^3} \log\left[\frac{(\omega^2_n+|\vec{p}|^2+\tilde{m}^2_1)...(\omega^2_n+|\vec{p}|^2+\tilde{m}^2_r)}{(\omega^2_n+|\vec{p}|^2+\tilde{m}^2_{r+1})...(\omega^2_n+|\vec{p}|^2+\tilde{m}^2_{r+q})}\right].
\end{equation}
In what follows, we limit ourselves to the case where effective potential originates from bosons\footnote{The extension to the fermionic case follows immediately, see \cref{pot_thermal}.}, for which $\omega_n= 2n \pi T$. After factoring out the common term $\alpha\equiv 1/(2\pi T)$ from the numerator and denominator, exchanging order of summation and integration and trading infinite sum for a product, we obtain
\begin{equation}
V^{\text{(gen)}}_{\text{eff}}= \frac{N_{\text{eff}}}{2}T\int \frac{d^3p}{(2\pi)^3} \log\left[\prod_{n\in\mathbb{Z}}\frac{(n^2+\alpha^2\tilde{E}^2_1)...(n^2+\alpha^2\tilde{E}^2_r)}{(n^2+\alpha^2\tilde{E}^2_{r+1})...(n^2+\alpha^2\tilde{E}^2_{r+q})}\right]+(\text{field-independent terms})\,.
 \label{V_gen_CH_prod_nr}
\end{equation}
Next, we make the order of polynomials in the numerator and denominator of \eqref{V_gen_CH_prod_nr} equal, by adding an auxiliary factor with some arbitrary field-independent constant $\tilde{E}^2_0$
\begin{equation}
V^{\text{(gen)}}_{\text{eff}}= \frac{N_{\text{eff}}}{2}T\int \frac{d^3p}{(2\pi)^3} \log\left[\prod_{n\in\mathbb{Z}}\frac{(n^2+\alpha^2\tilde{E}^2_0)^{r-q}(n^2+\alpha^2\tilde{E}^2_1)...(n^2+\alpha^2\tilde{E}^2_r)}{(n^2+\alpha^2\tilde{E}^2_{r+1})...(n^2+\alpha^2\tilde{E}^2_{r+q})}\right]+(\text{field-independent terms})\,.
 \label{V_gen_CH_prod}
\end{equation}
This transformation corresponds merely to an irrelevant shift of the effective potential by a constant independent of $\pi^{\hat{a}}$. The infinite product can be then separated to simple factors, and computed for each factor individually
\begin{equation}
\prod^\infty_{n=-\infty}\frac{n^2+\alpha^2\tilde{E}^2_i}{n^2+\alpha^2\tilde{E}^2_j}=\frac{\sinh^2(\pi\alpha^2\tilde{E}^2_i)}{\sinh^2(\pi\alpha^2\tilde{E}^2_j)}\,.
\end{equation}
After evaluating the products and dropping all the field-independent terms, we finally obtain
\begin{equation}
V^{\text{(gen)}}_{\text{eff}}=N_{\text{eff}}\sum^r_{i=1}\sum^{r+q}_{j=r+1}\int \frac{d^3p}{(2\pi)^3}\left[\frac{\tilde{E}_i}{2}+\log\left(1-e^{\tilde{E}_i/T}\right)-\frac{\tilde{E}_j}{2}-\log\left(1-e^{\tilde{E}_j/T}\right)\right].
\end{equation}
The above integral is a generalisation of \cref{eq:V^T_genII}. Formally, one can write it as
\begin{equation}
V^{\text{(gen)}}_{\text{eff}}=N_{\text{eff}}\sum^r_{i=1} \left[ V^{(T=0)}_{\text{CW}}(\tilde{m}_i)+\frac{T^4}{2\pi^2}J_B\left(\frac{\tilde{m}_i}{T} \right)\right]-N_{\text{eff}} \sum^{r+q}_{i=r+1}\left[ V^{(T=0)}_{\text{CW}}(\tilde{m}_i)+\frac{T^4}{2\pi^2}J_B\left(\frac{\tilde{m}_i}{T} \right)\right].
\end{equation}
We emphasize that cancellation of divergences in field-dependent terms of the potential is not a general feature of any composite Higgs models, unless some additional symmetries or sum rules are imposed. In most cases $V^{(T=0)}_{\text{CW}}$ diverges and its form depends on the choice of the regularisation scheme. For instance, in $\overline{\text{MS}}$ it reads~\cite{Dolan:1973qd}
\begin{equation}
V^{(T=0)}_{\text{CW}}(\tilde{m}^2_i)=\frac{\tilde{m}^4_i}{16\pi^2}\left[\log\left(\frac{\tilde{m}_i}{\mu}\right)-\frac{3}{2}\right].
\end{equation}
Whenever the potential is finite, the $\mu$ is just irrelevant auxiliary parameter which does not affect field-dependent part of the effective potential, see discussion below \cref{eq:VCWFULL}. In contrast, in the models where the effective potential must be regularised, $\mu$ takes a role of renormalization group scale and its choice affects physical predictions.

\section{Thermodynamics of phase transition}
\label{Thermo}

In this appendix we discuss the theoretical tools used to analyse the order and strength of the phase transition, conditions for bubble nucleation, and to estimate the GW spectra. We clearly state the approximations used in our analysis to obtain the predictions.

The phase transition is initiated by spontaneous nucleation of the true vacuum bubbles in the metastable phase. The probability per unit time and volume for tunneling between false and true vacuua at finite temperature is given by \cite{Linde:1980tt,Linde:1981zj}
\begin{equation}
\Gamma(T) = \left(\frac{S_3}{2\pi T}\right)^{\frac{3}{2}} T^4 \textrm{e}^{-S_3/T}, \label{vaccum_decay}
\end{equation}
where $S_3$ is the three dimensional Euclidean action. We compute the Euclidean action by using a modified version of the {\tt CosmoTransitions} code~\cite{Wainwright:2011kj}. The time at which bubbles of the true vacuum start to appear inside the false vacuum is given by the condition
\begin{equation}
    \label{cond1}
    \int_{t_{\rm c}}^{t_{\rm n}}\!{\rm d}t \frac{\Gamma(t)}{H(t)^3}=
    \int_{T_n}^{T_c}\!{\rm d}T \frac{\Gamma(T)}{H(t)^4 T} = 1.
  \end{equation}
where $T_c$, $T_n$ denote the critical and nucleation temperatures, respectively, and the Hubble expansion rate $H$ is given by
\begin{equation}
  H^2(T)=\frac{\rho_r}{3 M_{pl}^2}, \quad \rho_r = \frac{\pi^2}{30} g_{*}(T) T^4.
\end{equation}
Here, $\rho_r$ and $g_{*}$  denote the energy density of radiation and the number of degrees of freedom \cite{Saikawa:2018rcs}, respectively, while $M_{pl} = 2.4 \times 10^{18}$~GeV is the reduced Planck mass. We use ${\rm d}t = -a(T)/H(T)dT$ and the scale factor $a(T) \sim T^{-1}$ to convert integration variable to temperature.

If the relevant temperature scale is around the EW scale, as is the case in our scenario, the nucleation condition can be considerably simplified as 
\begin{equation}
    \frac{S_3}{T}\Big{|}_{T=T_n} \approx 140.
\end{equation}
The strength of the transition ($\alpha$) is quantified by the amount of latent heat released normalized to the radiation energy density. Here we take the strength to be defined as $1/4$ times the difference in trace of energy momentum tensor normalized to radiation, i.e.,
\begin{equation}
    \alpha \equiv  \frac{1}{\rho_r}\left( \Delta V(h,\eta,T) - \frac{T}{4} \Delta \frac{\partial V(h,\eta,T)}{\partial T} \right)\Big{|}_{T=T_n}, 
    \label{alpha_def}
\end{equation}
where the symbol $\Delta$ represents that the difference is taken between the false and true vacuum. The inverse time duration of the transition can be derived using \eqref{vaccum_decay} as
\begin{equation}
    \label{eq:betaH}
\frac{\beta}{H} \equiv T \frac{d}{dT} \left(  \frac{S_3}{T} \right)\Big{|}_{T=T_n}.
\end{equation}

Theoretical uncertainties in predicting the GW spectra from FOPT stem from estimating relevant thermodynamic parameters at the temperature when the PT concludes. For better accuracy, parameters should be evaluated at the percolation temperature ($T_p$), where the probability ($P$) of finding region of space in the false vacuum drops to about $P\sim e^{-1}$. However, to circumvent the computational bottleneck of evaluating $T_p$, we calculated thermodynamic parameters at the nucleation temperature ($T_n$). This is justified because the transitions in our scenario do not involve significant supercooling in the potential, implying that $T_p \approx T_n$. Another computational challenge involves in calculating the velocity at which the bubbles expand, for which we have used the following reasonable estimate, as shown in \cite{Lewicki:2021pgr}:

\begin{equation}
    \label{eqn:approx_velocity}
v_{w}\approx
\begin{cases}
\sqrt{\frac{\Delta V}{\alpha \rho_r}} \quad \quad {\rm for} \quad \sqrt{\frac{\Delta V}{\alpha \rho_r}}<v_J(\alpha),
\\
1 \quad \quad \quad \quad {\rm for} \quad  \sqrt{\frac{\Delta V}{\alpha \rho_r}} \geq v_J(\alpha) \, ,
\end{cases}
\end{equation}
where $v_J=\frac{1}{\sqrt{3}}\frac{1+\sqrt{3 \alpha^2+2 \alpha}}{1+\alpha} $ is the Chapman-Jouguet velocity which defines the upper limit for which hydrodynamic solutions can be found\footnote{Although the model discussed in this paper is more intricate than a simple extension of the SM with an elementary gauge singlet scalar, the use of \cref{eqn:approx_velocity} is justified, since at the transition scale effectively the same degrees of freedom contribute to the friction as in \cite{Lewicki:2021pgr}.}. 

\bibliography{main.bib}
\bibliographystyle{JHEP}
\end{document}